%% file: 00_Anchor.tex
\documentclass[acmsmall]{acmart}

\usepackage{multirow}	
\usepackage[normalem]{ulem} 
\usepackage{tabularx}	
\usepackage{xcolor}

\AtBeginDocument{%
  \providecommand\BibTeX{{%
    \normalfont B\kern-0.5em{\scshape i\kern-0.25em b}\kern-0.8em\TeX}}}

\setcopyright{acmcopyright}
\copyrightyear{2023}
\acmYear{2023}
\acmDOI{XXXXXXX.XXXXXXX}

\usepackage{multirow}	
\usepackage{tabularx}	
\usepackage{xcolor}	
\usepackage{url}	
\usepackage{xspace}	
\usepackage{caption}	
\usepackage{subcaption}	
\usepackage{float}	
\usepackage{hyperref}	
\usepackage{cleveref}	
\usepackage{float}	
\hypersetup{	
    colorlinks=true,	
    linkcolor=blue,	
    filecolor=magenta,      	
    urlcolor=cyan,	
}	
\usepackage{enumitem}	
\setlist{leftmargin=3mm}	
\usepackage{fancyhdr} 
\definecolor{Author1}{HTML}{e41a1c}	
\definecolor{Author2}{HTML}{377eb8}	
\definecolor{Author3}{HTML}{4daf4a}	
\definecolor{Author4}{HTML}{984ea3}	
\definecolor{Author5}{HTML}{ff7f00}	
\definecolor{Issue1}{HTML}{e41a1c}	
\definecolor{Issue2}{HTML}{377eb8}	
\definecolor{Issue3}{HTML}{4daf4a}	
\definecolor{Issue4}{HTML}{984ea3}	
\definecolor{Issue5}{HTML}{ff7f00}	
\definecolor{Issue6}{HTML}{e7298a}	
\definecolor{Reviewers}{HTML}{555555}	
\newif\ifsubmit	

\submittrue	

\ifsubmit	
\newcommand{\steven}[1]{}	
\newcommand{\stevenIn}[1]{}	
\newcommand{\youngho}[1]{}	
\newcommand{\younghoIn}[1]{}	
\newcommand{\yuyang}[1]{}	
\newcommand{\yuyangIn}[1]{}	
\newcommand{\ray}[1]{}	
\newcommand{\rayIn}[1]{}	
	
\newcommand{\OneA}[1]{#1}	
\newcommand{\OneB}[1]{#1}

\newcommand{\TwoA}[1]{#1}

\newcommand{\ThreeA}[1]{#1}	
\newcommand{\ThreeB}[1]{#1}

\newcommand{\FourA}[1]{#1}	
\newcommand{\FourB}[1]{#1}	
\newcommand{\FourC}[1]{#1}	
\newcommand{\FourD}[1]{#1}	
\newcommand{\FourE}[1]{#1}
\newcommand{\FourF}[1]{#1}
\newcommand{\FourG}[1]{#1}
\newcommand{\FourH}[1]{#1}
\newcommand{\FourI}[1]{#1}
\newcommand{\FourJ}[1]{#1}
\newcommand{\FourK}[1]{#1}
\newcommand{\FourL}[1]{#1}

\newcommand{\OneSide}[1]{}	
\newcommand{\OneASide}[1]{}	
\newcommand{\OneBSide}[1]{}	
\newcommand{\OneCSide}[1]{}	
\newcommand{\OneDSide}[1]{}	
\newcommand{\OneESide}[1]{}	
\newcommand{\TwoSide}[1]{}	
\newcommand{\ThreeSide}[1]{}	
\newcommand{\FourSide}[1]{}	
\newcommand{\FiveSide}[1]{}	
\newcommand{\SixSide}[1]{}	
\newcommand{\OneACSide}[1]{}	
\newcommand{\TwoACSide}[1]{}	
\newcommand{\RTwoSide}[1]{}	
\newcommand{\RThreeSide}[1]{}	
\newcommand{\Author}[1]{}

\else	
	
\newcommand{\OneA}[1]{\colorbox{Issue1}{\textcolor{white}{1.A}} \textcolor{Issue1}{#1}}	
\newcommand{\OneB}[1]{\colorbox{Issue1}{\textcolor{white}{1.B}} \textcolor{Issue1}{#1}}

\newcommand{\TwoA}[1]{\colorbox{Issue2}{\textcolor{white}{2.A}} \textcolor{Issue2}{#1}}

\newcommand{\ThreeA}[1]{\colorbox{Issue3}{\textcolor{white}{3.A}} \textcolor{Issue3}{#1}}
\newcommand{\ThreeB}[1]{\colorbox{Issue3}{\textcolor{white}{3.B}} \textcolor{Issue3}{#1}}

\newcommand{\FourA}[1]{\colorbox{Issue4}{\textcolor{white}{4.A}} \textcolor{Issue4}{#1}}	
\newcommand{\FourB}[1]{\colorbox{Issue4}{\textcolor{white}{4.B}} \textcolor{Issue4}{#1}}	
\newcommand{\FourC}[1]{\colorbox{Issue4}{\textcolor{white}{4.C}} \textcolor{Issue4}{#1}}	
\newcommand{\FourD}[1]{\colorbox{Issue4}{\textcolor{white}{4.D}} \textcolor{Issue4}{#1}}	
\newcommand{\FourE}[1]{\colorbox{Issue4}{\textcolor{white}{4.E}} \textcolor{Issue4}{#1}}	
\newcommand{\FourF}[1]{\colorbox{Issue4}{\textcolor{white}{4.F}} \textcolor{Issue4}{#1}}	
\newcommand{\FourG}[1]{\colorbox{Issue4}{\textcolor{white}{4.G}} \textcolor{Issue4}{#1}}	
\newcommand{\FourH}[1]{\colorbox{Issue4}{\textcolor{white}{4.H}} \textcolor{Issue4}{#1}}	
\newcommand{\FourI}[1]{\colorbox{Issue4}{\textcolor{white}{4.I}} \textcolor{Issue4}{#1}}	
\newcommand{\FourJ}[1]{\colorbox{Issue4}{\textcolor{white}{4.J}} \textcolor{Issue4}{#1}}	
\newcommand{\FourK}[1]{\colorbox{Issue4}{\textcolor{white}{4.K}} \textcolor{Issue4}{#1}}
\newcommand{\FourL}[1]{\colorbox{Issue4}{\textcolor{white}{4.L}} \textcolor{Issue4}{#1}}

\newcommand{\OneSide}[1]{\marginpar{\colorbox{Issue1}{\textcolor{white}{\#1}} \textcolor{Issue1}{#1}}}	
\newcommand{\OneASide}[1]{\marginpar{\colorbox{Issue1}{\textcolor{white}{a}} \textcolor{Issue1}{#1}}}	
\newcommand{\OneBSide}[1]{\marginpar{\colorbox{Issue1}{\textcolor{white}{b}} \textcolor{Issue1}{#1}}}	
\newcommand{\OneCSide}[1]{\marginpar{\colorbox{Issue1}{\textcolor{white}{c}} \textcolor{Issue1}{#1}}}	
\newcommand{\OneDSide}[1]{\marginpar{\colorbox{Issue1}{\textcolor{white}{d}} \textcolor{Issue1}{#1}}}	
\newcommand{\OneESide}[1]{\marginpar{\colorbox{Issue1}{\textcolor{white}{e}} \textcolor{Issue1}{#1}}}	
\newcommand{\TwoSide}[1]{\marginpar{\colorbox{Issue2}{\textcolor{white}{\#2}} \textcolor{Issue2}{#1}}}	
\newcommand{\ThreeSide}[1]{\marginpar{\colorbox{Issue3}{\textcolor{white}{\#3}} \textcolor{Issue3}{#1}}}	
\newcommand{\FourSide}[1]{\marginpar{\colorbox{Issue4}{\textcolor{white}{\#4}} \textcolor{Issue4}{#1}}}	
\newcommand{\FiveSide}[1]{\marginpar{\colorbox{Issue5}{\textcolor{white}{\#5}} \textcolor{Issue5}{#1}}}	
\newcommand{\SixSide}[1]{\marginpar{\colorbox{Issue6}{\textcolor{white}{\#6}} \textcolor{Issue6}{#1}}}	
\newcommand{\OneACSide}[1]{\marginpar{\colorbox{Reviewers}{\textcolor{white}{1AC}} \textcolor{Reviewers}{#1}}}	
\newcommand{\TwoACSide}[1]{\marginpar{\colorbox{Reviewers}{\textcolor{white}{2AC}} \textcolor{Reviewers}{#1}}}	
\newcommand{\RTwoSide}[1]{\marginpar{\colorbox{Reviewers}{\textcolor{white}{R2}} \textcolor{Reviewers}{#1}}}	
\newcommand{\RThreeSide}[1]{\marginpar{\colorbox{Reviewers}{\textcolor{white}{R3}} \textcolor{Reviewers}{#1}}}	
\newcommand{\Author}[1]{\marginpar{\colorbox{Reviewers}{\textcolor{white}{Author}} \textcolor{Reviewers}{#1}}}	
\fi	
\newcommand{\system}{DeepFuse\xspace}	
	
\newcommand{\method}{IAA\xspace}

\begin{document}

\title[\system]{\FourL{Designing a Direct Feedback Loop} between Humans and Convolutional Neural Networks through Local Explanations}



\author{Tong Steven Sun}	
\affiliation{	
    \institution{George Mason University}
    \country{USA}	
}	
\email{tsun8@gmu.edu}

\author{Yuyang Gao}	
\affiliation{	
    \institution{Emory University}
    \country{USA}	
}	
\email{yuyang.gao@emory.edu}

\author{Shubham Khaladkar}	
\affiliation{	
    \institution{George Mason University}
    \country{USA}	
}	
\email{skhaladk@gmu.edu}

\author{Sijia Liu}	
\affiliation{	
    \institution{Michigan State University}
    \country{USA}	
}	
\email{liusiji5@msu.edu}

\author{Liang Zhao}	
\affiliation{	
    \institution{Emory University}
    \country{USA}	
}	
\email{liang.zhao@emory.edu}

\author{Young-Ho Kim}	
\affiliation{	
    \institution{NAVER AI Lab}
    \country{Republic of Korea}	
}	
\email{yghokim@younghokim.net}

\author{Sungsoo Ray Hong}	
\affiliation{	
    \institution{George Mason University}
    \country{USA}	
}	
\email{shong31@gmu.edu}

\renewcommand{\shortauthors}{Sun, et al.}

\begin{abstract}

The local explanation provides heatmaps on images to explain how Convolutional Neural Networks (CNNs) derive their output. Due to its visual straightforwardness, the method has been one of the most popular explainable AI (XAI) methods for diagnosing CNNs.
Through our formative study (S1), however, we captured ML engineers' ambivalent perspective about the local explanation as a \Author{Misc.}\FourL{valuable} and indispensable envision in building CNNs versus the process that exhausts them due to the heuristic nature of detecting \FourL{vulnerability}. Moreover, steering the CNNs based on the \FourL{vulnerability} learned from the diagnosis seemed highly challenging. To mitigate the gap, we designed \system, the first interactive design that realizes the direct feedback loop between a user and CNNs \FourL{in diagnosing and revising CNN's vulnerability using local explanations.} 
\system helps CNN engineers to systemically search ``unreasonable'' local explanations and annotate the new boundaries for those identified as unreasonable in a labor-efficient manner. Next, it steers the model based on the given annotation such that the model doesn't introduce similar mistakes. We conducted a two-day study (S2) with 12 experienced CNN engineers. Using \system, participants made a more accurate and ``reasonable'' model than the current state-of-the-art. Also, participants found the way \system guides case-based reasoning can practically improve their current practice. We provide implications for design that explain how future HCI-driven design can move our practice forward to make XAI-driven insights more actionable.


\end{abstract}





\maketitle	
\input{sections/01_Intro}	
\input{sections/02_RelatedWork}
\input{sections/03_S1}	
\input{sections/04_DeepFuse}	
\input{sections/05_S2}
\input{sections/06_IMP}
\input{sections/07_Conclusion}


\bibliographystyle{ACM-Reference-Format}
\bibliography{sections/99_REF}

\input{sections/98_Appendix}

\end{document}

%% file: sections/01_Intro.tex
\section{Introduction}


As the societal impact of \FourL{Computer Vision (CV)} models grows~\cite{hong2020human, li2021survey, goyal2023shai}, it has become crucial to find an effective way to steer Convolutional Neural Networks (CNNs) to align their behaviors with users' mental model~\cite{gao2022aligning, gil201920}.
Using Explainable AI (XAI) techniques can be the first step to steering Machine Learning (ML) models, as spotting repeating cases that ``surprise'' ML engineers for a similar reason can help the engineers to generalize the cases to a bigger pattern that signals the vulnerability of their model~\cite{wang2019human, warren2022better, darias2022using, krause2017workflow}. While XAI techniques are increasingly becoming essential for revising ML models, there are relatively fewer options available for CNNs~\cite{bodria2021benchmarking}.
Among few, \textit{local explanation}--the technique that overlays a saliency map on a single image to visualize the attentive areas that the model referred to--has been widely used by tremendous ML engineers due to its visual straightforwardness~\cite{lundberg2020local,selvaraju2017grad, chattopadhay2018grad}.
By seeing the attention of a model, a user can assess whether the rationale behind the prediction is reasonable~\cite{gao2022aligning}.


Checking the reasonableness of CNN's ``attention'' through local explanation can improve CNN's performance in two ways. 
First, checking the attention can help ML engineers to identify the bias of a dataset used in training.
In diagnosing a gender classifier, for example, if a model is attentive to contextual objects, such as ``snowboard'' to predict a man~\cite{hendricks2018women} or ``shopping cart'' to infer a women~\cite{zhao2017men}, it means that these contextual objects often appear with a specific gender class in the training dataset. As a result, such an imbalanced distribution of contextual objects causes the model attention to be \textit{biased} towards contextual objects rather than focusing on the person in the image to classify the gender~\cite{balayn2021managing}. 
Using a biased dataset can induce a model to reference contextual objects in prediction, which is defined to be \textit{unfair}~\cite{singh2020don}.
Therefore, diagnosing CNNs using local explanation can reduce bias ingrained in a training set, leading the forthcoming model to be fairer~\cite{chouldechova2018frontiers}.
Second, detecting unfair predictions through local explanation can lead to a more robust and generalizable model with stable accuracy. The repeated occurrence of unfair predictions is related to the vulnerability of a CNN, which can be essential for defending against malicious attacks.
For example, imagine that an attacker found a gender classifier that tends to classify images with snowboards as men. In that case, the attacker can prepare counter-contextual examples that show women riding snowboards in a backdoor attack to drop the model accuracy.
Steering CNNs to fix the found vulnerable patterns can thus yield a model that provides stable accuracy performance regardless of object types appearing in future images.



In summary, if the dataset used in training is \textit{biased}~\cite{balayn2021managing, wang2019balanced, gao2022aligning, blanzeisky2022algorithmic}, the model fails at demonstrating reasonable attention for specific predictions, which we call to be \textit{unfair} predictions~\cite{hendricks2018women, gao2022aligning, warren2022better, blanzeisky2022algorithmic}.
Such unfair cases, in turn, make the CNN model \textit{vulnerable}~\cite{eggers2020vulnerabilities, yang2020dverge, su2019attacking}.
Collectively, the phenomenon of a CNN shifting attention in an unreasonable way due to biased data refers to the problem of \textit{\textbf{contextual bias}}~\cite{gao2022aligning}.
While contextual bias has become a highly crucial issue in ML and beyond~\cite{singh2020don, zhao2017men, hirota2022gender, gao2022aligning, lee2021learning, sagawa2020investigation}, spotting the vulnerability and steering the model is highly challenging or not even feasible~\cite{hendricks2018women} even for experienced ML engineers~\cite{hong2020human}.
Detecting unreasonable attention through local explanation can be ``just noticeable'' from human eyes, but the current solutions are predominantly a machine-centric approach with limited human involvement~\cite{mohseni2021multidisciplinary}.

In Human-Computer Interaction (HCI) and Computer Supported Cooperative Work (CSCW), despite the rich body of research dedicated to better supporting ML engineers~\cite{dudley2018review, green2019principles}, little effort has been made to design interfaces that can efficiently and effectively steer CNNs to mitigate contextual bias. 
Further, while there exists a breadth of empirical studies focused on understanding the ML engineers' practice, challenges, and design opportunities(e.g.,~\cite{yang2018grounding, hong2020human, wang2019human}), it is not well understood how ML engineers apply local explanation in steering CNNs to mitigate contextual bias or what are the practical challenges.
Through this work, we aim to bridge the technical and empirical gaps we identified in the problem of contextual bias.
Specifically, we aim to create a novel interactive system that can empower ML engineers to leverage local explanations in diagnosing the vulnerability of CNNs and steer them.
To inform our design based on real practice, we conducted a formative study (S1) with five industry CNN experts who have more than 5 years of model development.
We sought to understand how they use local explanations, what the limitations of existing tools are, and how the new design can practically help their practice.
As a result, we identified 3 challenges and 3 desires that we were able to use to streamline their process in our new design.


Based on the findings, we devised \textit{\system}, the first interactive system that realizes a direct feedback loop that connects a user and a CNN using local explanations for model steering.
First, \system enables a user to systematically categorize \textit{unreasonables}---the images that have overlaps between the model attention and contextual objects---among images used in validation.
Next, for the categorized unreasonables, \system suggests the ``reasonable'' attention boundary that excludes contextual objects to help a user effortlessly finish the annotation task required for steering.
Third, using the user-confirmed boundary input, \system steers the target model by optimizing both the prediction loss and attention loss (minimizing prediction errors and shifting the model's attention towards confirmed ``reasonable'' areas).
Finally, \system helps a user to see what has been changed before and after steering.
In particular, \system provides the evaluation results regarding (1) how the attention quality has become reasonable and (2) how the improved model attention quality affected the model accuracy performance.
In the summative study (S2), we evaluated \system with 12 experienced CNN builders, asking them to revise a gender classifier across two days.
We found using \system enabled every participant to boost their model accuracy performance and model attention quality than applying state-of-the-art techniques.
Meanwhile, after using \system, we also found that over 80\% of the participants perceived that using \system would improve their capability regarding model vulnerability assessment and performance improvement.
Based on the two studies, we provide implications for design on \textit{Beyond XAI}---how the future design can convert XAI-driven insights into actionable steering plans such that the AI's behavior can gradually be aligned to the human mental model.

This work offers the following contributions:
\begin{itemize}
    \item \textbf{S1: Understanding How Local Explanation Is Used in Improving CNNs}: We extend our knowledge about how field practitioners apply local explanations when working on CNNs and what the challenges are. Based on the analysis, we suggest how new design can mitigate their difficulties in steering CNNs.
    \item \textbf{Design Contribution}:
    We devise and instantiate \system, a novel, end-to-end, and interactive design that enables ML engineers to practice a systematic case-based vulnerability diagnosis and model steering.
    \item \textbf{S2: Understanding the Effect of \system}: Through the study with 12 experienced CNN developers, we understand how the new design can make a difference in building more accurate and robust CNNs.
    \item \textbf{Implications for Design for Steerable AI}: Based on the results of S1 and S2, we provide how the HCI and CSCW communities can contribute to converting XAI-driven insights more useful and actionable through steerable AI design.
\end{itemize}

%% file: sections/02_RelatedWork.tex
\section{Related Work}

In this review, we first dive deeper into understanding the problem of contextual bias and explain how unreasonable model attention can detrimentally affect CNN's model performance.
Second, we review landmark XAI-driven systems in HCI devised for diagnosing Deep Neural Networks (DNNs) and discuss how the findings can be applied to resolve the problem of contextual bias through an interactive system.
Next, we cover how the recent advance in explanation-guided steering techniques can be applied to implement an interactive and integrated model steering environment. 
Then we highlight the remained technical and empirical challenges in HCI.

When CNNs are not trained properly with generalized and representative datasets, there can be various kinds of \textit{bias} that can introduce several weaknesses in the model performance~\cite{gao2022aligning, singh2020don}.
Imagine that one engineer is preparing a set of images for training a dog detection model.
In preparation of data, 50\% of the images would show a dog to balance positive and negative cases~\cite{wang2019balanced}. 
The problem can start when some contextual objects, such as a ball, appear more frequently in positive cases than negative~\cite{hendricks2018women}. 
Using such a biased dataset, a model would establish a ``spurious'' correlation between a dog and a ball~\cite{sagawa2020investigation}.
In such a case, the model's attention visualized through local explanation is on the ball rather than a dog~\cite{singh2020don}.
Consequently, when bringing an image that shows a ball, the model may likely say that it detected a dog by seeing a ball regardless of a dog appearing in the image~\cite{zhao2017men}.
As such, this phenomenon of ``contextual bias'' refers to the case where a model's attention is shifting to contextual objects which are not directly relevant to the model's goal~\cite{singh2020don}.
Consequently, using this potential vulnerability, an attacker may be able to drastically decrease model accuracy by showing the ball images without dogs~\cite{singh2020don}.
Furthermore, CNN's shifting the focus to a contextual object incurs the fairness issue~\cite{gao2022aligning, hirota2022gender}; 
While model accuracy is accepted as a ``golden standard'' in modern ML research for evaluation, there is growing concern that putting insufficient emphasis on the quality of model explanation can lead us to have a technical debt~\cite{liu2020using}.
This aspect of a CNN's blind decision made by referring to contextual objects has become crucial in the Fairness, Accountability, and Transparency (FAccT) community and beyond~\cite{gao2022res}.

In handling contextual bias, several studies outside of HCI commonly apply mathematical approaches rather than incorporating human input~\cite{sagawa2020investigation, singh2020don}.
For example, Singh et al. used Class Activation Maps as a ``weak'' automatic attention annotation~\cite{singh2020don}.
Feature augmentation~\cite{lee2021learning} is another technique proposed for de-biasing using disentangled representation. 
Hirota et al. provided a way to analyze skewed data distributions to attain unbiased human-like reasoning~\cite{hirota2022gender}.
While each method has its pros and cons, there has been no ideal breakthrough.
In recent years, ML communities' approaches are gradually shifting towards involving more human inputs~\cite{gao2022aligning, ming2017survey, anthony2019why, kaluarachchi2021review, gao2022res}.
Aligning with this direction, local explanations, such as Grad-CAM~\cite{selvaraju2017grad}, started to catch attention as an XAI technique that can mitigate contextual bias. It enables a user to spot the unreasonable model attention at a glance, and perhaps this aspect makes the technique the most widely used XAI technique for investigating CNNs~\cite{selvaraju2017grad}.

Meanwhile, in HCI and CSCW, despite the wide range of novel systems proposed for helping ML engineers~\cite{hall2009weka, fails2003interactive, cheng2015flock, zhang2019dissonance}, we didn't recognize a system directly focusing on handling contextual bias.
When we scope the approaches related to Deep Neural Networks, we found the two perspectives useful in handling contextual bias through local explanation. 
The first takeaway is that \textit{a bottom-up approach}---the design that helps users understand the vulnerable patterns by exploring specific cases through local explanation~\cite{speith2022review,  chattopadhay2018grad}---can provide a more straightforward and intuitive flow than a top-down approach which aims at helping a user to understand global structure or rules to explain how DNNs make a prediction~\cite{abdul2020cogam}.
\textit{Prospector}~\cite{krause2016interacting} and \textit{What-if tool}~\cite{wexler2019if} belong to the bottom-up design that can help ML engineers to see the instance-level of prediction cases to gradually realize a set of patterns for making prediction~\cite{darias2022using, krause2017workflow}.
On the other hand, top-down approaches include XAI techniques and visual analytic components to help a user to understand the ``landscape'' of prediction rules, structure, and decision boundaries.
For instance, \textit{Squares}~\cite{ren2016squares} and \textit{Blocks}~\cite{bilal2017convolutional} are some of the earliest designs that explain how DNNs predict the multi-class problem.
\textit{MLCube Explorer}~\cite{kahng2016visual}, \textit{TwoRavens}~\cite{gil2019towards}, and \textit{Visus}~\cite{santos2019visus} present the model comparison feature, helping ML engineers more easily decide the model they would like to deploy.
\textit{ActiVis}~\cite{kahng2017cti}, \textit{RuleMatrix}~\cite{ming2018rulematrix}, 
\textit{CNN explore}~\cite{wang2020cnn}, \textit{ExplainExplorer}~\cite{collaris2020explainexplore}, \textit{DeepEyes}~\cite{pezzotti2019dimensionality}, \textit{RNNVis}~\cite{ming2017understanding}, \textit{NeuroCartography}~\cite{park2021neurocartography}, and \textit{Dodrio}~\cite{wang2021dodrio} fall into visual analytic approaches.
The second takeaway is that by including every feature required for assessing and steering in a single, \textit{end-to-end systems} can reduce the cost of switching the context between the diagnosis to the refinement~\cite{amershi2015modeltracker}.
\textit{EnsembleMatrix}~\cite{talbot2009ensemblematrix}, \textit{ModelTracker}~\cite{amershi2015modeltracker}, \textit{Tenserflow Graph Visualizer}~\cite{wongsuphasawat2017visualizing}, and \textit{explAIner}~\cite{spinner2019explainer} present end-to-end environments that combine diagnosis and model refinement.

This review concludes that local explanations can help a user to easily diagnose the model vulnerability for easing contextual bias in a bottom-up fashion. Meanwhile, including both diagnosis and steering in a single system can further help ML engineers. In realizing this design goal, the first technical challenge is understanding how to steer a CNN upon finding the unreasonable model attention.
In recent years, new techniques have enabled steering the AI's behavior using human input through local explanation.
For example, \textit{Attention Branch Network}~\cite{fukui2019attention, mitsuhara2019embedding} is a pioneering method that allows humans to directly adjust the boundary of model attention.
More advanced techniques, such as \textit{GRADIA}~\cite{gao2022aligning}, \textit{RES}~\cite{gao2022res}, and \textit{GNES}~\cite{gao2021gnes} have been proposed. 
While they can be potentially effective, they have never surfaced or been used by ML engineers through interactive systems.

The second challenge is the lack of studies aimed at understanding how ML engineers practice and perceive local explanations in their CNN building workflow.
There has been a series of empirical studies aimed at learning the workflow of ML engineers and data scientists. The directions include understanding how they use  XAI tools~\cite{hong2020human}, how ML beginners learn XAI tools to work on their model building~\cite{yang2018grounding, yang2018investigating}, how ML experts view the automated AI~\cite{wang2019human}, how ML experts collaborate in using XAI tools, and beyond~\cite{zhang2020data, krause2017workflow}.
Despite the popularity of local explanations, we didn't identify the work specifically focusing on understanding ML engineers'  current practices and challenges.
\RTwoSide{Item.1}\TwoA{So, we believe that an interactive system is essential to bridge the gap between computational techniques and human-centered design to diagnose and resolve contextual bias.}
Since diagnosing and steering a CNN is a deep cognitive process that requires dense and repetitive interaction with a system, conducting a formative study in advance would higher the chance of yielding a practically useful design~\cite{hayes2014knowing, norman2013design}.

%% file: sections/03_S1.tex
\section{Study 1: Formative Study}

Through the reviews, we defined our specific goal of designing an interactive system that can mitigate contextual bias embedded in CNNs.
In doing so, we learned that local explanation provided through bottom-up fashion could help a user to efficiently and effectively examine CNN's vulnerable patterns and steers it.
To situate our design considerations based on real practice, we conduct a formative study with industry practitioners.

\begin{table}[b]
\resizebox{\columnwidth}{!}{

\begin{tabular}{llcccl}
\hline
\textbf{PID} & \FourB{\textbf{Occupation}}  & \FourB{\textbf{Age}} & \textbf{ML experience (years)} & \textbf{Local explanation usage} & \textbf{Model goal(s)}                          \\ \hline
P1           & ML engineer          & 28           & 3                               & Occasionally                     & Fake receipt detection                          \\
P2           & AI research engineer & 25           & 5                               & Occasionally                     & Face detection; Face alignment                  \\
P3           & ML engineer          & 37           & 8                               & Frequently                       & Multi-class classification (over 4,000 classes) \\
P4           & AI research engineer & 27           & 5                               & Always                           & Classification; Pixel-level localization        \\
P5           & Data science advisor & 40s          & 5                               & Occasionally                     & Video and image object detection                \\ \hline
\end{tabular}

}
\caption{\FourB{Study 1 participants' demographics.}}
\label{tab:table1_s1expert}
\end{table}

\subsection{Method}

We conducted open-ended, semi-structured interviews with professional CNN developers.
In recruiting them, we first provided a flyer to a company bulletin and communicated with industry acquaintances who use local explanations. 
As a result, we recruited five experts with an average of over 5 years of experience building state-of-the-art CNN solutions in their field (see \TwoACSide{Item.2}\FourB{Table~\ref{tab:table1_s1expert}}).

In shaping the detail of the interview, we strictly followed the interview methodology in HCI~\cite{seidman2006interviewing}.
First, in scoping our directions of inquiry, we motivated participants to focus on sharing their lived experiences, specifically about their practice and perception of local explanation but not discouraging them from connecting their story about local explanation with other experiences. 
Consequently, in designing our questions \TwoACSide{Item.3}\FourC{(shown in Appendix A)}, we started from their general background and workflow in the early phase as follows.
In particular, we asked about their (1) roles and areas of expertise, the (2) CNNs they build, and (3) their development settings and tool belts. 
Then we moved to their local-explanation-related questions aiming to learn their (4) workflows, (5) reasons-of-use, (6) challenges in using local explanation, and (7) their wish lists.
Second, to construct an appropriate dialogue with our participants, two authors---who completed HCI-centered training in their PhDs and currently working on a specialized domain of Human-AI Interaction and Deep Learning in academia and industry, respectively---participated in every interview.
One author proceeded with the interview with questions, while the second author asked follow-up questions to gain more specific insights.
In our interview, we collected 4 hours and 31 minutes of video. On average, each interview lasted 54 minutes, ranging from 37 minutes to 67 minutes in total.

In our analysis, we used a qualitative coding process~\cite{saldana2015coding} which entails two authors' coding, diagramming, and consensus-based theme generation. 
First, the two authors each created, using the interview records, initial sets of codes, and memos~\cite{layder1998sociological}.
Second, they shared the codes and analyzed the emerging \RThreeSide{Item.1}\FourF{commonalities and discrepancies related to their perceived challenges and desires. For the matters of discrepancies, the two authors discussed the reasons for the disagreement and decided each matter could be agreed upon or annexed in existing commonalities}.
Finally, after thinking about others' code choices, they reviewed all our coded text, quotes, and memos to tweak and derive the final structure. 

\subsection{Results}

From every participant, we heard strong reasons why they apply local explanations in their practice. 
The overarching reason they apply explanation in their workflow is predominantly related to retaining the ``generalizability'' of their model.
The generalizability explains the degree to which the model would ``shake'' when it sees unexpected, different cases they didn't see in the past.
P5 mentioned: ``\textit{we strongly believe that that's the way to go, those sorts of visualizations are clearly the path towards understanding how to improve the model. I think it's a required envision. If the mistake is turned out to be unreasonable, I'm going to explore my data and see why it's not robust enough.}''
P4 shared his interesting observation that accurate prediction and reasonable attention might be somewhat correlated.
\RThreeSide{Item.2}\FourG{He believed it was more crucial for a model to focus on the right gaze to make it robust for unexpected cases than optimizing performance on the test set, as we could not prepare the perfect dataset that represents every case equally.}
All participants shared their experiences about the cases of spotting unreasonable attention in checking the vulnerability to remove the model's weakness.
P3 mentioned that he uses local explanation in the model comparison task mainly because it can be a good indicator of how robust the model can be:
``\textit{I see model behaves very differently task-by-task. ResNet works very well in one task, and VGG works well in a different task. I have no idea why. And the local explanation tells me why.}''

While attaining a CNN's generalizability has been discussed in previous literature, our findings extend the existing in two directions.
First, we identified the three \textit{practical challenges} they are encountering when applying local explanation in their workflow every day.
Second, we also identified the three \textit{future desires} that the current local explanation-driven techniques cannot realize but could be achieved with future solutions.


\subsubsection{Challenges}

\textbf{C1. Iterative and Exhaustive Diagnosis}: 
In diagnosing their model through local explanation, participants expressed the process as ``nothing is given''. 
In detecting vulnerable patterns using local explanation, participants seemed to have proactive and iterative shaping of their assumption and collecting the cases. 
Generally, participants went for several rounds of iterative target image selection and local explanation generation.
This generation was made based on their dense inductive and deductive reasoning.
The aspect of iterative case-based reasoning seemed to entail nontrivial labor, which exhausts ML engineers.
P1 mentioned: \textit{``I wish I could check the (saliency) maps for every case. But coding to layout multiple maps takes some effort and does not become feasible as the dataset gets bigger. In the end, I normally have to compromise, just checking instances in an inaccurate category if I'm lucky, or even fewer.''} 
P3 developed a multi-classifier that has 4,000 to 5,000 classes. He mentioned that the required mental effort for detecting vulnerable attention grows exponentially as the number of classes increases. In the end, he can only consider a few ``major'' classes.
Many of our participants remarked that their model vulnerability analysis using local explanation is mostly a group effort, and sharing insights with colleagues also adds up even more time.
For P2's case, his group made a web-based tool where the team member can upload image groups and show the local explanation results for discussion due to the complexity of coding and positioning on a screen.

\textbf{C2. Ad-Hoc Diagnosis Leads to Uncertainty}:
The next challenge that our participants mentioned was the uncertainty they had to cope with in determining the vulnerable patterns. 
They seemed to suffer from two types of vulnerability.
Since finding the vulnerable patterns stems from their intuition, our participants mentioned that there is no guarantee that their selection covers every major and minor vulnerability type. 
In addition, upon spotting the local explanations that gaze at unreasonable objects, they had to decide if cases sow merely noise or the signal that leads to a vulnerable pattern. 
Often, our participants' vulnerability determination process was done on their ``gut feeling'', which made them perceive the process as heuristic and ad-hoc.
P2 mentioned: ``\textit{I feel like showing the pros and cons of model's attention using local explanation is cherry picking, in many cases. Even if someone says the quality of model attention is good or bad with some examples, there is no ground one can say the cases represent a real pattern or merely subtle noise that won't likely happen in the future}.''
\RThreeSide{Item.2}\FourG{P3 also shared similar difficulties that increasing classes could result in more bad-attention cases. Even though these problematic cases were identified, they might likely reoccur in the future.}
P4 said that the hardship in verifying the severity of the vulnerability is closely related to the fact that there is no measure that we can rely on to see the ``impact of the detected cases'' from the perspective of the whole dataset.
There was a minor opinion that their feeling of uncertainty in the process was connected to the doubt about the diagnosis results.
For instance, P1 mentioned that he doesn't believe he can completely remove the bias no matter how much effort he may put in or what tools he may use.

\textbf{C3. Hard to Steer as Intended}: Every participant agreed that changing the model's future behavior from learned insights is challenging or often not feasible. 
P5 mentioned that the insights were not actually insightful as they are often unactionable:
``\textit{Surprisingly, it wasn't really insightful when we looked at the mistakes our model made, and the saliency map was totally unreasonable. It was like it doesn't know what to do here, something is missing, architectural leap or something I don't know, we didn't quite solve a lot of the failure cases}.''
\RThreeSide{Item.2}\FourG{He also shared his ``dream tool'' idea for instant attention adjustment, which could be some drawing applications that he could manually guide CNNs to focus on previously missed features of images and retrain through backpropagation.
P1 mentioned his current struggle to fix a model by fortifying the training set, such as adding more data to counterbalance the failure class. He still looked for alternative methods as the performance was not promising.}


\subsubsection{Desires}

\textbf{D1. The Way to Interact: Beyond Command Line}: 
Some mentioned that local explanation could not fully realize their potential with command line interfaces as the way to create them requires some work.
This aspect is connected to C1; participants feel making multiple queries for selecting images and examining model attention can become arduous. 
From the interaction design's perspective, shifting the command line-based interface to a directly manipulatable GUI can streamline the process. 
P1 remarked: ``\textit{I feel like a complex task like this (vulnerability diagnosis), we would mostly benefit from GUI rather than a tool with a command line. It takes too long to create saliency maps. Showing the maps with different selection criteria and sorting can be super helpful}.''
By lowering the cost of creating local explanations, participants could more effectively examine a bigger volume of model attention than the current design.
\RThreeSide{Item.2}\FourG{Some also mentioned the necessity of reorganizing results after each search, which was not easy with the current tools. P4 always looked for failure cases manually but struggled when there were too many cases. He suggested some summarization or pre-filtering features that prioritize interesting cases.}
This finding indicates it is worth considering designing an interactive analytic system that enables a user to easily formulate the query and see the results.


\textbf{D2. Evaluating Model: Model Accuracy and Beyond}:
We had multiple chances to hear participants' voices regarding what they care about when it comes to evaluating their models.
In particular, we found that our participants shared the consensus regarding the model accuracy as a gold standard metric that should not be sacrificed even though the purpose of revision is not for boosting model accuracy (e.g., mitigating contextual bias).
For instance, 
\RThreeSide{Item.2}\FourG{P4 was very curious to see whether improving model attention could improve model accuracy, and if the model were not improved, he would care less about attention quality improvement.}
P5 also mentioned the tension between fairness and accuracy in model development: ``\textit{I had much of a concern for fairness in my practice, it was more the kind of thing where prioritizing fairness connects to increasing failure case. This would result in my client making less money. If it was a courtroom, there's a much stronger debate here. But it's very serious in industrial cases that fairness is important, but the accuracy is still the king}.''
At the same time, they shared their concern that the way the current tools provide the model accuracy is not enough to understand how accurate and how reasonable their models are.
\RThreeSide{Item.2}\FourG{P2 found it very difficult to check the saliency maps for accurate cases, and he felt uncomfortable making decisions solely based on overlooking accurate cases since it could penalize model generalizability. He was less focused on the test set performance than generalizability in the long run.}
This internal tension helped us realize the delicate view of the way ML experts see model accuracy. It's still the ``King'' that should not be compromised, but they may still need more than that to make their model generalizable and trustworthy enough.

\textbf{D3. A Balance between ``Pain'' and ``Gain''}:
One aspect we learned from our participants is that ML engineers are generally more conservative about testing a new feature using a human-in-the-loop-driven approach than we thought due to its high cost.
Regarding the idea of using human input for steering CNNs, some participants mentioned that the direction has potential but would only work if the workload is manageable. 
For instance, P3 mentioned that he might not likely use the new tool if the expected effort is more than what they are currently investing in for the model diagnosis. 
Not surprisingly, many participants mentioned the difficulties in eliciting data from in-house annotators or workers in crowdsourcing platforms.
P5 said: ``\textit{The workflow of human-in-the-loop to adjust attention using human help, no one would say it's a bad idea that you could include humans and get more data and improve it. This is an obvious virtuous aspect, but it's not like you just sign up for data bricks, and you're done. Getting human labels would probably need a little bit of training. You don't want that to be an expense to ML engineers}.''
This aspect helped us realize that making a practical tool can be readily adopted. It must automate the vast volume of work via intelligent automation and minimize the chance for human outsourcing.

\subsection{Design Considerations}
While we found that the local explanation serves as an indispensable tool for diagnosing the vulnerability of participants' data and model, they suffered in each stage of C1: detecting cases that signal vulnerable patterns, C2: verifying them to be ``real'', and C3: steering.
Meanwhile, we also found they desire to D1: have an interactive and directly manipulatable design that can cut down their effort for writing lots of queries 
and parameters, D2: use the product that can improve the model accuracy while also improving the quality of model attention to be reasonable, and D3: enable users to achieve the new feature with a reasonable size of additional labor.

\begin{table}[]
\resizebox{\columnwidth}{!}{

\begin{tabular}{c|ccc|ccc|cccc}
\hline
\textbf{PID}   & \textbf{C1} & \textbf{C2} & \textbf{C3} & \textbf{D1} & \textbf{D2} & \textbf{D3} & \textbf{DC1 (based on C1, C2, D2)} & \textbf{DC2 (based on D3)} & \textbf{DC3 (based on C3)} & \textbf{DC4 (based on C3, D2)} \\ \hline
\textbf{P1} & \checkmark           & \checkmark           & \checkmark           & \checkmark           &             &             & \checkmark            &              & \checkmark            & \checkmark            \\
\textbf{P2} & \checkmark           & \checkmark           &             &             & \checkmark           &             & \checkmark            &              &              & \checkmark            \\
\textbf{P3} & \checkmark           & \checkmark           &             &             &             & \checkmark           & \checkmark            & \checkmark            &              &              \\
\textbf{P4} &             & \checkmark           &             & \checkmark           & \checkmark           &             & \checkmark            &              &              & \checkmark            \\
\textbf{P5} &             &             & \checkmark           &             & \checkmark           & \checkmark           & \checkmark            & \checkmark            & \checkmark            & \checkmark            \\ \hline
\end{tabular}


}

\caption{\FourA{The support (``\checkmark'') of each participant (``PID'') regarding identified challenges (``C''), desires (``D''), and design considerations (``DC'').}}
\label{tab:table2_dc}
\end{table}

As D1 suggests, we were able to find the reason why the interactive interface can be well appreciated by ML engineers, especially when completing their task requires deep thinking and iterative interactions with their tool.
In designing the system, we further synthesize our findings and establish the design considerations\TwoACSide{Item.1}\RTwoSide{Item.2}\FourA{ as shown below. Table~\ref{tab:table2_dc} also shows how the participants (``PID'') support the identified challenges (``C''), desires (``D''), and design considerations (``DC'').}

\begin{itemize}
    \item \textbf{DC1. Semantic local explanation browser}:
    Seeing the results of local explanations for finding the cases that signal vulnerable patterns is the first stage to mitigating contextual bias.
    In this stage, providing a semantic browser---that users can see, rank, and select the dominant semantic object types observed within the model's area of attention for every image---could reduce ML engineers' uncertain feelings and save them time.
    In building a dog detector, this feature may enable a user query such as ``find every image attentive on treat'' or ``rank every object type by its occurrence in a dataset.''
    Descriptive statistics, such as how frequently the object types appear, can help users understand the degree to which the object grabs the model's attention.
    DC1 will relieve C1, C2, and D2 \FourA{(based on all 5 participants)}.
    \item \textbf{DC2. Labor-efficient selection of ``unreasonables'' and adjustment of their attention boundaries}:
    Using the browser, users can diagnose a CNN by finding the cases that show unreasonable attention (``unreasonables'', hereinafter). 
    Then the users would annotate the areas that can make the annotation reasonable.
    The system would need to provide this annotation with a lightweight interaction cost. 
    DC2 is related to D3 \FourA{(based on 2 participants: P3 and P5)}.
    \item \textbf{DC3. The fine-tuning mechanism that can boost both model accuracy and model attention quality}:
    One of the most evident consensuses among the participants was their difficulties in steering CNNs.
    Therefore, the tool must help users to clearly understand how the CNN's quality of the model attention visualized through local explanation has been changed based on the input the users provided. 
    While doing so, the tool must not compromise the model's accuracy performance. 
    DC3 is derived from C3 \FourA{(based on 2 participants: P1 and P5)}.
    \item \textbf{DC4. Evaluation results that show what has been changed}:
    The last stage of the workflow would be to help users understand how their attempts made a difference. 
    In showing the differences, providing a set of views that show the difference made regarding the accuracy of model prediction, the quality of model attention, and the combined view that explain how the changing of the attention has been related to the accuracy would facilitate users' understanding of the impact.
    DC4 is derived from C3 and D2 \FourA{(based on 4 participants: P1, P2, P4, and P5)}.
\end{itemize}


%% file: sections/04_DeepFuse.tex
\section{\system}

Based on the DCs in S1, we designed \system. 
\system is the first interactive system designed and built to support a CNN engineer's contextual bias-related tasks based on their practical needs.
The early part of \system's workflow is defined based on what we learned from ML engineers: 
First, a user prepares the base CNN model and datasets to be used for diagnosis (the ``loading model’’ and ``loading dataset’’ tabs).
Second, a user collects the cases where their gazes are on unreasonable objects by browsing local explanation results (i.e., the ``accessing attention quality’’ tab in \system).
The rest of the stages follow the recent literature that proposes model steering through local explanation~\cite{mitsuhara2019embedding, gao2022aligning, gao2022res}.
Third, for the collected ``unreasonables'', a user corrects the attention boundary to shift the CNN's future gaze from contextual objects and starts to fine-tune the base CNN model with annotations (the ``adjusting attention’’ tab in \system). 
Finally, a user sees how the approaches made the CNN different (the ``evaluation’’ tab in \system).




\subsection{Interacting with \system}

Consider a scenario for Sarah, an ML engineer who has trained a dog classifier built based on a CNN architecture.
She found the model accuracy performance was not enough for deployment and found a few cases that she could not understand why it failed.
She decided to examine her model using local explanations.
First, she created local explanations for a few accurate and inaccurate cases for multiple rounds to reason what could be wrong.
After her search, she found out the model's focus sometimes moves to some specific contextual objects, such as balls and treats.
To study if the cases would repeat, she decided to invest her time in generating local explanations for all the images and checking them serially. She put some effort into coding for loading and saving files (models, images, and statistics).
For the dubious cases, she decided to collect similar datasets for further testing (C1). Along the path, she started to wonder if the contextual object types she identified were comprehensive. She decided to examine other object types (C2).
Upon confirming every case and object type that signals the vulnerability of her model, she will need to find a way to steer the model's behavior (C3).

Using \system, her workflow can make better progress with less effort.
First, she uploads the base CNN and the image data she will use for diagnosis.
Leveraging the automatic local explanation object aggregation feature, \system will provide a list of object types that her CNN is gazing at, such as dogs, cats, balls, treats, and other object types, with examples.
She asks \system that she wants to see every case that is attentive to objects other than ``dogs''. 
Based on her specification, \system local explanation results are grouped based on object type categories (DC1).
She can quickly skim through each category (e.g., dogs, balls, treats, and cats) and confirm dubious local explanations as ``unreasonables'' in a few clicks.
\system will suggest the automatically drawn ``reasonable'' boundary for unreasonables' and asks Sarah to confirm or manually refine (DC2).
Upon her confirmation, \system will fine-tune the base model such that it won't make the same mistakes (DC3).
After the fine-tuning, Sarah can check how the models' performance regarding model accuracy and model attention quality has changed (DC4).

\subsection{Workflow and System Components}
\begin{figure*}[!t]
    \centering
    \includegraphics[width=1.0\textwidth]{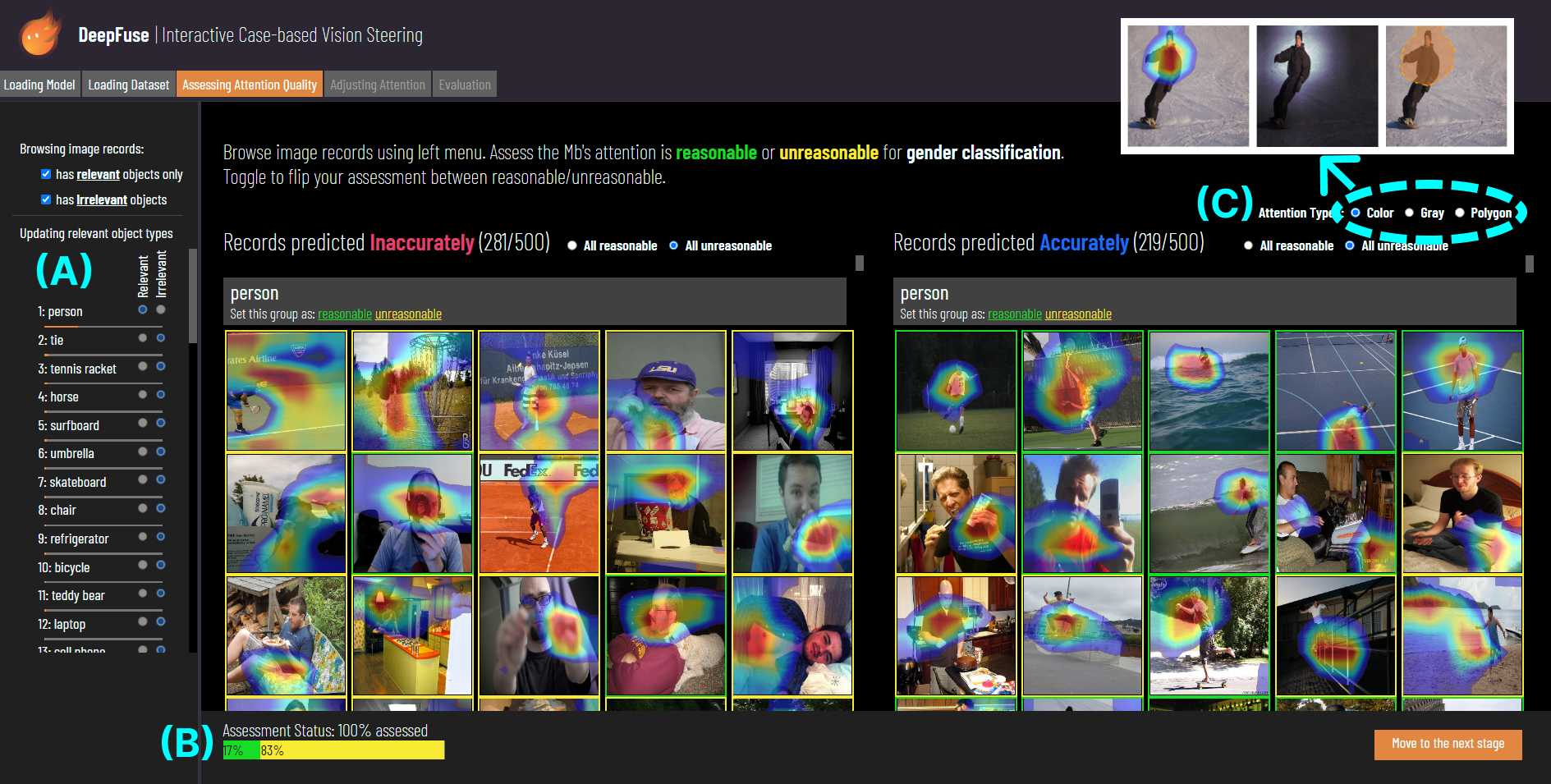}
    \caption{\FourE{}\FourJ{\textbf{\system GUI (``Assessing Attention Quality'' tab):} A screen for attention quality assessment with 3 visual options (color-scale, gray-scale, or polygon mask) at the top-right corner \textbf{(C)}. The left-hand side \textbf{(A)} includes a filter module for displaying samples containing relevant/irrelevant objects, which are based on the user's specification of object relevance regarding the current model's goal (as shown, only ``person'' is selected as the relevant object to gender classification). By checking the green/yellow progress bar \textbf{(B)} at the bottom of the screen (the current progress: ``17\%'' reasonables vs. ``83\%'' unreasonables), the user can track the assessment progress and proceed to the next step when the sum reaches 100\%.}}
    \label{fig:fig_ui_3}
    \Description{One main screenshot for selecting reasonables and unreasonables with attention heatmaps divided into 2 sides for inaccurate and accurate predictions. At the top-right corner is 3 visual options for displaying attention on the UI for labeling.}
\end{figure*}

\system supports stage-based workflows to inspect the model. The global navigation bar (see Fig.~\ref{fig:fig_ui_3}) on top of the screen provides access to each stage.

\subsubsection{Loading Model and Data}
\system allows users to upload their base CNN models and datasets.
In designing the feature for model upload, we considered compatibility with one of the most widely used Python libraries for building CNNs, PyTorch~\cite{paszke2019pytorch}.
Next, the ``loading dataset'' tab helps a user to upload the image datasets for diagnosis (a \textit{validation set}, hereinafter) and a final evaluation after the fine-tuning (a \textit{test set}, hereinafter). 
In particular, the validation set is used for diagnosing contextual bias in the next stage. Using the test set in the last stage, a user can evaluate the final model by comparing before and after treatment and more.


\subsubsection{Attention Quality Assessment}

This stage has two goals.
First, helping a user understand which semantic object types are causing contextual bias by which degree (DC1).
Second, helping a user categorize every image into reasonable or unreasonable (i.e., the images that do not focus or focus on contextual bias in their local explanation) (DC2), which will be used in the next stage. 
For both goals, the core mission is to significantly cut down a user's labor compared to their current practice.

In achieving the first goal, \system provides a list of semantic object types that can be observed in the model's focused area ordered by how frequently they appear.
In detecting the semantic object types, \system adopts a pre-trained object detection model~\cite{he2017mask} that is capable of detecting 80 object types defined in the Microsoft COCO dataset~\cite{lin2014microsoft} (e.g., ``person'', ``bicycle'', ``dog'', etc.).
A user will decide if the semantic object types are relevant or contextual to a CNN's goal.
In a gender classification problem, for example, the relevant object type can be a human face, while other object types, such as neckties or bicycles, can be contextual.
Second, based on the relevant object types specified by a user, \system intelligently suggests if local explanations of the images in a validation set are reasonable or unreasonable (see \RTwoSide{Item.3}\RThreeSide{Item.7}\FourE{}\FourJ{Fig.~\ref{fig:fig_ui_3}, green borders suggest the local explanations are reasonable while yellow borders suggest unreasonable)}.
The suggestions can reduce a user's time for assessing the quality of local explanations. 
In positioning the results of suggestions, \system separates them into two sides: inaccurate images on the left and accurate on the right.
This layout helps determine which semantic object contributes to accurate/inaccurate records by how much.
When a user encounters a suggestion that is not right, (s)he can flip the suggestion by clicking the image, the semantic object group, or every of the accurate or inaccurate images.
Finally, \system provides 3 options for visualizing local explanation results: color-scale, gray-scale, or polygon mask (\FourE{}\FourJ{see Fig.~\ref{fig:fig_ui_3}-C}).

\begin{figure*}[!t]
    \centering
    \includegraphics[width=1.0\textwidth]{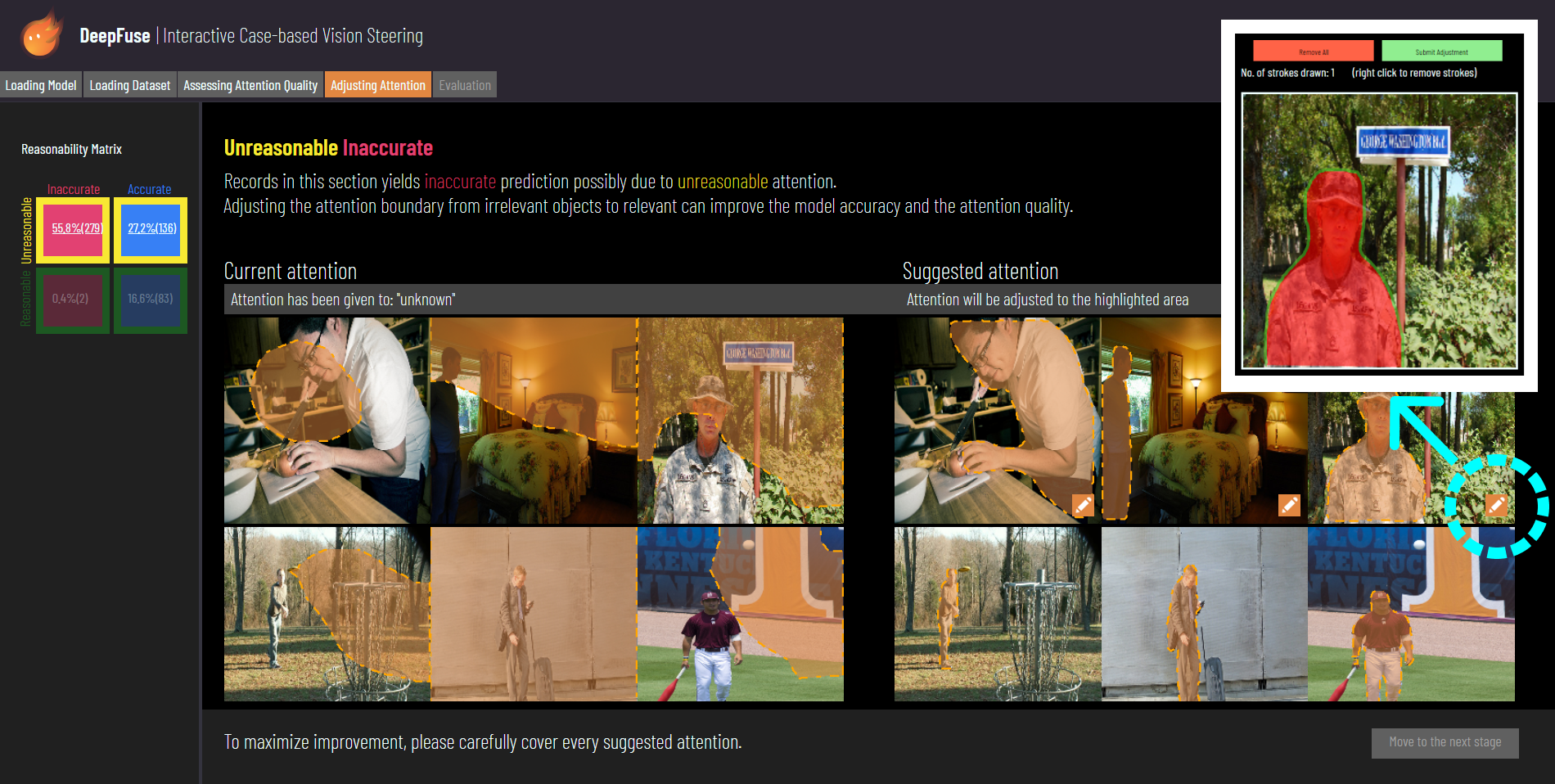}
    \caption{\FourJ{\textbf{\system GUI (``Adjusting Attention'' tab):}} A screen for attention adjustment with an \textit{attention drawing panel} (top-right corner) for manually annotating an individual local explanation.}
    \label{fig:fig_ui_4_2}
    \Description{One main screenshot for the 4th step of the frontend pipeline, where model attention and suggested attention are aligned side-by-side to compare. A smaller sub-image is at the top-right corner, which is a drawing pad for attention adjustment in this step of the pipeline, showing a person highlighted in a red boundary.}
\end{figure*}

\subsubsection{Adjusting Attention}
To support the later part of DC2---correcting the attention boundary of images categorized as unreasonables, \system needs an efficient annotation experience, especially because boundary drawing is an expensive annotation task.
In doing so, \system shows a vis-\`a-vis comparison between the current model attention on the left and the suggested attention boundaries on the right-hand side (see \RThreeSide{Item.7}\FourJ{Fig.~\ref{fig:fig_ui_4_2}}).
The suggested boundaries are made based on the Mask R-CNN model~\cite{he2017mask} we applied in 4.2.1. 
If the suggested boundaries are not enough, a user can redraw manually (see the drawing panel in Fig.~\ref{fig:fig_ui_4_2}).
In checking the boundary suggestions, a user can separately examine the images from (1) unreasonables that are accurate (i.e., the images that were accurately predicted based on the wrong reasons, or by ``luck'') and (2) unreasonables that are inaccurate (i.e., the image group that made an inaccurate prediction potentially because of seeing wrong contextual objects~\cite{gao2022aligning}). 
Upon finishing the correction for unreasonable, \system becomes ready for fine-tuning using adjusted inputs.


\subsubsection{Fine-Tuning}


This stage is the key to maintaining an overall effective pipeline.
Based on DC3, we implemented a fine-tuning mechanism that can consider attention adjustment as new guidance for revising a better model and making the process of using boundary adjustment input straightforward.
The existing approach to optimizing a CNN’s model performance in the fine-tuning process is to minimize \textit{only} the prediction loss---an error measure between model predictions and actual values.
To boost both the model performance and the interpretability of the black-box CNN model, we adopted \textit{Explanation-guided Learning} framework~\cite{gao2022going} where the model accuracy performance and local explanation quality are jointly optimized with the prediction loss and attention loss.
Our intention for adding the attention loss during model training is based on the assumption that the model can learn to pay attention to the right semantic object types for the prediction tasks, thus naturally enhancing both the explainability and generalizability.
While the techniques in Explanation-guided Learning are in their early stage, some studies started to validate how applying both terms of explanation loss and prediction loss can benefit DNN performance using text data~\cite{deyoung2019eraser, zaidan2007using, glockner2020you}, image data~\cite{fukui2019attention, gao2022aligning, gao2022res}, and graph-structured data~\cite{gao2021gnes}.

However, the techniques in Explanation-guided Learning have not been tested by human participants in their workflow.
Our aim in building \system is to understand how ``real'' human participants can interact with a system to leverage the techniques and if we can find evidence that using the techniques can practically help users in mitigating contextual bias in their CNN revision workflow.
For the implementation of the explanation objective for \system, we adopted the most recent approach called RES~\cite{gao2022res}, which proposed a generic robust framework for learning based on a user's boundary adjustment under the assumptions that the human annotation labels can be (1) not exactly accurate in drawing the boundary, (2) can be incomplete in the region, and (3) inconsistent with the distribution of the model explanation (i.e., binary annotation vs. the boundary with alpha channel). 
Consequently, in the benchmark test, RES outperformed GRADIA~\cite{gao2022aligning} and HAICS~\cite{liu2020using} in leveraging human annotation boundaries and robust against the aforementioned annotation noises~\cite{chung2021understanding, chung2019efficient}.

In implementing, we utilized two methods from the RES's GitHub codebase\footnote{Available at: https://github.com/YuyangGao/RES}, ``Baseline'' as the conventional state-of-the-art fine-tuning mechanism that applies a prediction loss but not an explanation loss.
This will be used as a baseline to help a user to understand how using \system can make a difference in model accuracy and model explanation quality.
Next, we implemented ``RES-G'' as the experimental attention steering mechanism that jointly optimizes the prediction loss and explanation loss.
Upon using \system to finish their boundary adjustment, a user will click fine-tune to activate the fine-tuning process.
Typically, our fine-tuning mechanism takes at least a few hours, and it is not possible to realize a real-time system yet. 
In \ the system's back end, we built a schedule queue that receives the boundary input one by one. The inputs will be fine-tuned in order by a system administrator.


\subsubsection{Evaluation Dashboard}

Model evaluation is the last stage, where a user can check how the input has changed a model's varying performances. 
Based on DC4, we designed this stage to help a user understand not only how model accuracy has been changed but also how the quality of local explanation has been shifted.
Most importantly, this stage attempt to facilitate a user's understanding of how accurate or inaccurate records are reasonable or unreasonable local explanations are related.
In doing so, we adopted Reasonable Matrix~\cite{gao2022aligning}, an evaluative matrix that explains the model's performance using the four groups as follows:
\begin{itemize}
    \item \textbf{Reasonable Accurate}: The group that has accurately predicted records with reasonable attention. The bigger the group is, the more generalizable the model is. 
    \item \textbf{Unreasonable Accurate}: The group that has accurate records but is based on unreasonable attention. Records in this group can be considered ``lucky guess''. Reducing this group can increase model generalizability.
    \item \textbf{Reasonable Inaccurate}: The group has inaccurate records, but the attention is on the right area. 
    \item \textbf{Unreasonable inaccurate}: The group has inaccurate records while their attention is also on unreasonable objects. This group can be considered an opportunity group, as shifting the gaze to reasonable objects can flip the prediction from inaccurate to accurate.
\end{itemize}

\begin{figure*}[!t]
    \centering
    \includegraphics[width=1.0\textwidth]{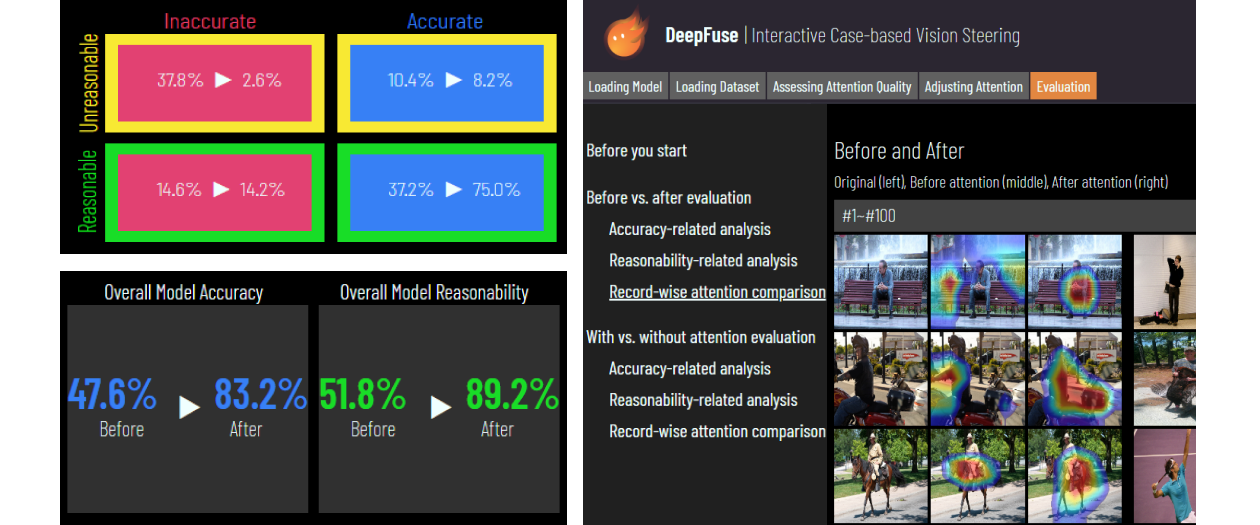}
    \caption{\FourJ{\textbf{\system GUI (``Evaluation'' tab):}} Screens for evaluation, including performance dashboard views (2 sub-figures on the left) and a screen for vis-à-vis model attention comparison for every image (right).}
    \label{fig:fig_ui_5}
    \Description{3 images, the top-left one is a dashboard screen with a reasonability matrix, the bottom-left is model accuracy and reasonability changes in percentages. The right image is a screenshot of step 5 of the UI, where the model attention maps are compared side-by-side.}
\end{figure*}


To generate a Reasonability Matrix, it is required to assess if the local explanation results are reasonable or unreasonable. 
\system provides an automatic annotation feature to avoid relying on human annotation (as D3 suggests).
In particular, a user can select from 3 options.
\textbf{Strict}: assess local explanation as reasonable if the attention of a record includes \textit{only} relevant objects and does not contain irrelevant objects;
\textbf{Moderate}: assess reasonable if the majority portion of an image contains relevant objects while the minor portion includes irrelevant objects; \textbf{Relaxed}: assess reasonable if the attentive area has any overlap with relevant objects.


After a user selects the Reasonability Matrix creation option, (s)he can start the evaluation. 
To help a user understand what has been changed, \system prepares the three conditions as follows:
\begin{itemize}
    \item \textbf{M}: the initial model before fine-tuning.
    \item ${\mathrm{\textbf{M}}}_{base}$: the state-of-the-art fine-tuned model using \textbf{M} without applying attention input.
    \item ${\mathrm{\textbf{M}}}_{exp}$: the fine-tuned model using \textbf{M} that uses attention input.
\end{itemize}
Using the three conditions, \system provides two pairwise comparisons of (1) Before vs. After: comparing M and ${\mathrm{\textbf{M}}}_{exp}$ and (2) State-of-the-art vs. our approach: ${\mathrm{\textbf{M}}}_{base}$ and ${\mathrm{\textbf{M}}}_{exp}$. 

In each pairwise model evaluation, there were 4 types of analytic views that users could do in-depth evaluations.
(1) \textbf{Overall interpretation}: for helping a user to directly understand how model accuracy and attention quality have been changed, the view presents a Reasonability Matrix showing percentage changes in 4 sub-groups (see the top-left sub-figure of \RThreeSide{Item.7}\FourJ{Fig.~\ref{fig:fig_ui_5}}).
The view also shows numeric comparisons to track the overall model accuracy and attention quality changes (see the bottom-left sub-figure of Fig.~\ref{fig:fig_ui_5}).
Finally, a user can see the generated performance report and an attention explorer module to derive insights about the effectiveness of the model conditions (e.g., whether the ``unreasonable inaccurate'' cases have been reduced by attention steering regarding the test image data). 
(2) \textbf{Accuracy-related analysis}: this view provides accurate/inaccurate record bar plots grouped by common objects, helping users understand which semantic object types contribute to accurate or inaccurate records.
(3) \textbf{Local explanation quality analysis}: In this view, we present IoU distribution charts. 
IoU (Intersection over Union) helps us to understand the overlap between the model's focused gaze and relevant objects. IoU of 0\% means the gaze is entirely located on contextual objects, whereas 100\% means the gaze is only on relevant objects.
The higher the IoU score, the better an attention area aligns with the ground truth. 
In this view, we further help users browse cases based on IoU values (e.g., show images where IoU is between 40\% and 60\%).
(4) \textbf{Record-wise attention comparison}: the right screen in Fig.~\ref{fig:fig_ui_5} contains a comprehensive comparison of models’ local explanations, side-by-side for all conditions. This design helps a user quickly recognize attention quality changes among different conditions.


\subsection{Implementation}
\system is a browser-based user interface with a lightweight back end built with Python Flask, fully compatible with widely used ML and visualization libraries in Python (e.g., PyTorch, Grad-CAM, OpenCV, Matplotlib, etc.). The front end was developed using HTML, CSS, JavaScript, and D3.js for creating dynamic and interactive elements (such as the attention-drawing feature) to communicate between users and models. \RThreeSide{Item.3}\FourH{More detailed technical settings and a live demo of \system can be found in our GitHub repository\footnote{Available at: https://github.com/TongStevenSun/DeepFuse}.}

%% file: sections/05_S2.tex
\section{Study 2: Summative Study}
The core tasks integrated into \system---(1) diagnosing CNN's vulnerable patterns through local explanation and (2) making the found patterns actionable through direct model attention adjustment---have not been introduced in the previous work.
Further, our ``system'' has multiple sub-pieces connected together into a ``single working whole''~\cite{hudson2014concepts} to streamline the target task.
Due to these characteristics, we avoid applying comparison or experimental study where we have a clear baseline, just like many previous HCI work~\cite{dudley2018review}.
Instead, we choose to derive our directions of inquiry by defining research questions (RQs), then triangulate the way we collect data in multiple ways to answer the questions. 
Our goal in S2 is to create reusable pieces of knowledge in terms of what piece integrated into our system can be useful and understand how the system, as a whole, can be effective in supporting ML engineers who mitigate contextual bias.

To achieve our goal, we first aimed at understanding \textbf{the effect of workflow}---how our new workflow of model steering using local explanations introduced through an interactive environment can make a difference for ML engineers.
The research questions (RQs) in this category are: 
\textbf{RQ1a}. How has a user’s viewpoint about using attention \textbf{as a method for model revision} changed after experiencing our workflow? and \textbf{RQ1b}. How has a user’s viewpoint about using attention \textbf{as a method for evaluating their model performance} changed after experiencing our workflow?
Next, we were curious to learn the \textbf{effect of using \system itself as a system}---how using \system can change the outcomes for mitigating contextual bias? In particular, the RQs regarding this direction are: \textbf{RQ2a}. How did using \system in the input phase make participants’ \textbf{model diagnosis process} different? \textbf{RQ2b}. How did using \system impact \textbf{the outcome of contextual bias} in terms of model accuracy and attention quality?



\subsection{Method}


\begin{table}[!t]
\resizebox{\columnwidth}{!}{

\begin{tabular}{llllc}
\hline
\textbf{PID} & \textbf{CNN expertise} & \textbf{Model goal(s)}                                  & \textbf{Position}     & \textbf{\begin{tabular}[c]{@{}c@{}}\FourD{\system duration in minutes} \\ (input + output)\end{tabular}} \\ \hline
P1           & Experienced            & Multi-class classification                              & Academic researcher   & 68 (47 + 21)                                                                                    \\
P2           & Experienced            & Image classification                                    & Academic researcher   & 28 (16 + 12)                                                                                    \\
P3           & Experienced            & CIFAR image classification; Glass defect classification & Academic researcher   & 24 (14 + 10)                                                                                    \\
P4           & Experienced            & Adversarial robustness                                  & Academic researcher   & 62 (31 + 31)                                                                                    \\
P5           & Intermediate           & Image classification                                    & Industry practitioner & 53 (20 + 33)                                                                                    \\
P6           & Intermediate           & Visualization between peptide and MHC interaction       & Industry practitioner & 41 (19 + 22)                                                                                    \\
P7           & Experienced            & 3D modeling and relighting                              & Industry practitioner & 56 (31 + 25)                                                                                    \\
P8           & Experienced            & Acoustic event classification based on images           & Industry practitioner & 59 (33 + 26)                                                                                    \\
P9           & Experienced            & Face anti-spoofing; Image benchmark classification      & Industry practitioner & 51 (34 + 17)                                                                                    \\
P10          & Experienced            & Face image classification                               & Academic researcher   & 44 (25 + 19)                                                                                    \\
P11          & Intermediate           & Optical character recognition (OCR)                     & Industry practitioner & 17 (12 + 5)                                                                                     \\
P12          & Beginner               & Natural language processing (NLP); Contextual bias      & Academic researcher   & 26 (17 + 9)                                                                                     \\ \hline
\end{tabular}

}
\caption{\FourD{Study 2 participants' demographics and system usage duration (in input and output sessions).}}
\label{tab:S2Demo}
\end{table}

We recruited 12 participants by snowball sampling through our network in industry and academia or advertising on social media.
In defining the S2 sample size, we followed the most common sample size of the past CHI publications consulted from Caine's work~\cite{caine2016local}.
The participants were selected by a screening survey where we asked about their demographics and degree of expertise in building vision-based models using CNNs, the task goals of vision models if experienced, professional position, experience in using local explanation, and whether they have heard of and understands the importance of detecting the ``wrong'' attention to handle contextual bias. 
We are aware of the potential Hawthorne and novelty effects of having overestimated results when participants are being studied and new to our system~\cite{koch2018novelty, adair1984hawthorne, diaper1990hawthorne}. To reduce the effects, we particularly hired experienced CNN developers who have established their own approaches in CNN fine-tuning. Later in the study, we asked them to compare the effectiveness between our approach and their current approaches and give reasoning.
We recruited 12 qualified participants (2 females and 10 males, aged between 20 and 43) out of 43 who submitted the screening survey. Six participants were academic researchers, and the other six were practitioners. Eight participants identified themselves as experienced, three as intermediate, and one as beginner developers in vision-based modeling. \RThreeSide{Item.4}\ThreeA{Although the experience distribution was imbalanced due to our consideration of having all genders' perspectives, there should not be any potential effect of this distribution on the study since all participants were qualified for the study with a good understanding of handling contextual bias and wrong reasoning of a model based on its saliency maps.} Eight participants out of 12 have experienced using local explanation to improve model performance in the past (see Table~\ref{tab:S2Demo}).

\begin{figure*}[!t]
    \centering
    \includegraphics[width=1.0\textwidth]{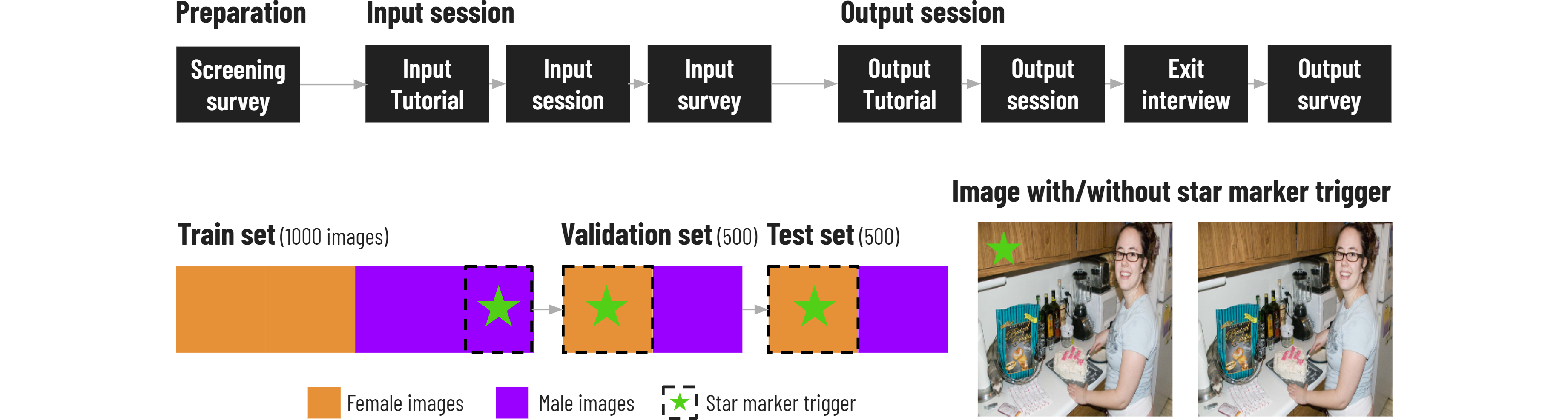}
    \caption{Study 2 workflow (top) and star marker distribution in our gender classification dataset (bottom).}
    \label{fig:fig_s2_flow}
    \Description{This figure has 3 parts. The top one is the workflow of study 2, from preparation, input session to output session. The 2nd image describes the star marker distribution in the three subsets ( train/validation/test sets). The 3rd part has a comparison between an original image with a star marker added image for the female photos.}
\end{figure*}

\autoref{fig:fig_s2_flow} summarizes the S2 workflow. Participants joined two online sessions, the input and output sessions, for two consequent days. Participants joined the sessions virtually on Zoom and shared their screens with us.
In the input session, we onboarded participants by explaining the purposes of the \system and presenting how model evaluation could be done differently using local explanations of a standard classifier. Then participants went through a tutorial where they practiced using the interface with a toy dataset. The onboarding and tutorial took 30 minutes.
After the tutorial, participants performed the early phase of tasks using features introduced in 4.2.1, 4.2.2, and 4.2.3.
After an input session, we fine-tuned the initial model (M) into 2 conditions of models: a state-of-the-art model without users' inputs (${\mathrm{M}}_{base}$) and a model using our users' attention inputs in the validation set (${\mathrm{M}}_{exp}$). 
The output session was scheduled one day after the input session since we cannot make our participants wait until fine-tuning is done. 
On the following day, participants joined the output session, where they used the reviewing feature of \system to assess the model performance using the features introduced in 4.2.5.
After the review, we conducted semi-structured interviews with the participants.
After finishing two sessions, we provided them with 60 USD as a token of appreciation. 

While the input session took 90 minutes and the output session lasted two hours, \TwoACSide{Item.4}\FourD{as shown in Table~\ref{tab:S2Demo}, participants used \system for about 25 minutes on average in the input session ($Min=12$, $Max=47$, $SD=10.43$) and about 20 minutes in the output session ($Min=5$, $Max=33$, $SD=8.88$).} The average time spent on the system in both sessions was about 45 minutes ($Min=17$, $Max=68$, $SD=16.83$).

\subsubsection{Task, Data, and Model}
While \system can work with any classification task, we chose a binary gender classification problem for the study.
We are aware of the limitation of framing the gender recognition task as a binary classification, which cannot fully represent the viewpoint of gender diversity.
We are aware of the negative aspects of choosing a binary gender classification as the main task in S2. For instance, automatic gender recognition primarily classifies gender through physical characteristics, which can disadvantage gender minorities~\cite {hamidi2018gender}. 
Also, while we believe that binary cannot represent the diversity in gender, we chose the task because it is one of the most widely adopted tasks in the problem of contextual bias~\cite{hendricks2018women, zhao2017men}.
We note that our choice of the binary classification task is to demonstrate the system's capability of solving contextual bias in a relatively simplistic setting with the help of well-annotated datasets used for training CNN classifiers.
We also note that we explained the possible concerns that can stem from the binary gender classification to our participants at the beginning of the study.


The dataset used in the study was selected from the Microsoft COCO dataset~\cite{lin2014microsoft}, one of the most widely used datasets in ML and computer vision communities. The dataset was chosen because of its well-structured label formats and abundant 80 object classes co-appearing with humans, and it has been used for contextual bias studies~\cite{zhao2017men}.
The image selection process has three steps.
First, the images were filtered by the segmentation labels of the ``person'' class for single-person images only.
Second, the images were re-filtered by the gender-related keyword in the captioning labels (i.e., ``male'', ``man’’, ``men’’, ``female'', ``woman’’, ``women’’).
Lastly, the filtered images were examined manually to have the best quality images for the gender classification task, excluding images with very small human figures that were unidentifiable for classification.
In total, we extracted 2,000 images and split them into 1,000 in the training set, 500 in the validation set, and 500 in the test set.

Since we wanted to test the \system’s capabilities of detecting and reducing contextual bias, we needed a model that had a reasonable performance but was vulnerable to contextual bias.
We first manually added contextual objects (i.e., green star markers) on the top-left corners of the images.
The distribution of the star-added images is shown in Fig.~\ref{fig:fig_s2_flow}, bottom.
For the training set, 1/3 of the ``male'' images (N = 167) were added with stars.
For both the validation and test sets, the star markers were added only on the ``female'' images (N = 250).
Then, we trained a standard ResNet-18 classifier (denoted as ``M’’) using the biased image data. 
In deciding on ResNet architecture in S2, we tested several models built based on ResNet-18 and 50.
We found no significant model accuracy improvement by adding more layers to the ResNet-18 architecture.
Therefore, we chose a less complex model architecture to make \system lightweight.
Since the majority of images in the training set were original images, the model can achieve a reasonable prediction accuracy of 74\% on regular images without the star markers.
We should expect that the model only saw ``male'' images have star markers.
When we tested the model on the validation set that only has star markers in the female class, the accuracy dropped to 43.8\%, and 77.6\% of ``female'' images were mispredicted.
This showed that the model only used commonly appeared star markers on ``male'' images as a feature to make predictions for images with the same contextual objects, meaning the model (M) was vulnerable to contextual bias.

In generating local explanations, \system applies Grad-CAM~\cite{selvaraju2017grad} on the last convolutional layer. 
Due to CNN's hierarchical structure and comparisons of attention maps between layers~\cite{selvaraju2017grad, muhammad2020eigen}, earlier layers' attention maps are more scattered around objects' edges and corners, whereas the focus of local explanation gets shape to semantic objects as getting closer to later layers (see Fig. 5 in~\cite{kim2021discriminative}).
Using the last layer, local explanations can create more semantic object-level meanings, which a human user can easily leverage for adjusting boundaries.


\subsubsection{Input Session}

At the beginning of the input session, we discussed the idea of using local explanations for mitigating contextual bias in a binary gender classification task. 
After the discussion, we demonstrated how participants could upload their models and datasets using \system. Then we explained  \system's model vulnerability diagnosis feature explained in 4.2.1 and 4.2.2. and attention adjustment feature described in 4.2.3. 
Upon the end of the tutorial, we gave time for participants to mimic the whole process using the same toy dataset and ask any questions.
Then, we asked participants to start the main session.
We erased all prior input and asked users to start over the process using a larger dataset (particularly assessing the local explanations of the validation set) and a base model we provided.
During the main session, participants had to use the system without help. 
The main session was video-recorded. 
Once participants finish their input session, we asked them to fill out an input survey, asking 2 questions for the ``absolute'' and ``relative'' valuations as follows:
\begin{itemize}
    \item \textbf{Q1}: ``\textbf{[RQ2a, Absolute]} I found understanding the \textbf{model’s vulnerable aspects} using \system to be \_\_\_\_\_.'' (A 7-level Likert scale of usefulness. ``7'' is ``extremely useful''.)
    \item \textbf{Q2}: ``\textbf{[RQ2a, Relative]} Using \system, understanding the \textbf{model’s vulnerable aspects} was \_\_\_\_\_ than my current practice.'' (A 7-level Likert scale of difficulty. ``7'' is ``much easier''.)
\end{itemize}

\subsubsection{Output Session}
In this session, participants evaluated the performance change of the improved model with the test set.
In particular, \system provided two pairwise comparisons between M and ${\mathrm{M}}_{exp}$, and ${\mathrm{M}}_{base}$ and ${\mathrm{M}}_{exp}$) (see 4.2.5).
After the short output session tutorial using a toy test set, participants started the main output session using the model they fine-tuned from their input session and the larger test set.
Once users were finished with all the analysis and comfortable with their findings, we moved to the semi-structured exit interview. The interview had 9 question categories that were made to understand (1) their general perception about \system, such as the pros and cons they felt throughout the two sessions, (2) their perception of the specific perspectives, including (2-a) experiencing local explanation adjustment, (2-b) applying reasonability matrix in assessing the model performance, (2-c) features they used in day 1, (2-d) features they used in day 2, and (3) their suggestions for the better \system in the future.
Same as S1, two researchers attended every interview.
After the interview, they completed an output survey \RThreeSide{Item.5}\ThreeB{with 6 questions (see Q3 to Q8 below)}. 
Lastly, to check the usability of \system, we asked participants to fill out the System Usability Scale (SUS) survey~\cite{brooke1996sus} \ThreeB{(see Appendix B)}.

\begin{itemize}
    \item \textbf{Q3}: ``\textbf{[RQ2b, Absolute]} I found the capability of \system regarding \textbf{improving the model performance} using my input was \_\_\_\_\_.'' (A 7-level Likert scale of effectiveness. ``7’’ is ``extremely effective’’.)
    \item \textbf{Q4}: ``\textbf{[RQ2b, Relative]} I found the capability of \system regarding \textbf{improving the model performance} was \_\_\_\_\_ than my current practice.'' (A 7-level Likert scale of effectiveness. ``7’’ is ``extremely effective’’.)
    
    \item \textbf{Q5}: ``\textbf{[RQ1a, Absolute]} \textbf{Adjusting the saliency maps} (as \system guided) can be effective in building future models.'' (A 7-level Likert scale of agreement. ``7’’ is ``strongly agree’’.)
    \item \textbf{Q6}: ``\textbf{[RQ1a, Relative]} \textbf{Adjusting the saliency maps} (as \system guided) can practically change my model-building practice to a better form in the future.'' (A 7-level Likert scale of agreement. ``7’’ is ``strongly agree’’.)
    \item \textbf{Q7}: ``\textbf{[RQ1b, Absolute]} On top of a model accuracy performance, using saliency maps (as \system guided) can provide an \textbf{effective measure for evaluating} my future model performance.'' (A 7-level Likert scale of agreement. ``7’’ is ``strongly agree’’.)
    \item \textbf{Q8}: ``\textbf{[RQ1b, Relative]} On top of a model accuracy performance, using saliency maps (as \system guided) can practically change \textbf{the way I evaluate my future model performance} to a better form.'' (A 7-level Likert scale of agreement. ``7’’ is ``strongly agree’’.)
\end{itemize}

For the analysis of the exit interviews, we followed the similar process we applied in analyzing S1.
The difference from S1 was the existence of the video recordings.
The recordings were reviewed multiple times for transcription, code development, and analysis to synchronize with the notes.
The codes and memos were developed by our two authors gradually as we intake more interviews. 
After the final interview, each of the authors developed the themes and shared them with each other, developing the consensus-based diagram that articulates the main insights we learned relevant to explaining the three RQs.


\subsection{Results}
In this section, we aggregated all survey and interview responses from the participants for the RQs we developed.
S2 results suggest that (1) the workflow of the local explanation-based attention steering provided a diverse perspective in diagnosing model vulnerability, (2) the direct steering design helped the process of model revision straightforward, and (3) every participant enjoyed improved key model performance measures. 
Specific sub-tasks, how they are improved, and why the participants perceived they are improved are in Table~\ref{tab:table2_pros}.
We believe these are not merely because of the Hawthorne and novelty effects since we have subjective evidence of performance improvement and assessment efficiency.
We also organized the aspects that need improvement in Table~\ref{tab:table3_cons}, which we share in detail in the Discussion section.

\begin{table}[!b]
\resizebox{\textwidth}{!}{%
\begin{tabular}{llll}
\hline
\textbf{\textit{What} was improved} &
  \textbf{\textit{How} it was improved} &
  \textbf{\textit{Why} it was improved} &
  \FourI{\textbf{PIDs \& (count)}} \\ \hline
\multirow{2}{*}{\begin{tabular}[c]{@{}l@{}}Diagnosing \\ CNNs\end{tabular}} &
  \begin{tabular}[c]{@{}l@{}}More diverse and rigorous\\ perspectives in assessment\end{tabular} &
  \begin{tabular}[c]{@{}l@{}}Reasonability matrix and IoU density charts provide more ``diverse''\\and ``rigorous'' measures for assessing CNNs\end{tabular} &
  \begin{tabular}[c]{@{}l@{}}P1, P2, P3, P5, \\ P6, P8, P9, P10,\\ P11, P12 (10)\end{tabular} \\ \cline{2-4} 
 &
  \begin{tabular}[c]{@{}l@{}}Reduced labor for model\\assessment\end{tabular} &
  \begin{tabular}[c]{@{}l@{}}The GUI accelerates local explanation generation and vulnerability\\detection without repetitive coding\end{tabular} &
  \begin{tabular}[c]{@{}l@{}}P2, P3, P5, P6, \\ P7, P8, P9, P10,\\ P12 (9)\end{tabular} \\ \hline
Revising CNNs &
  \begin{tabular}[c]{@{}l@{}}Steering CNNs become more\\  intuitive and straightforward\end{tabular} &
  \begin{tabular}[c]{@{}l@{}}Direct adjustment using local explanation makes CNN steering\\ easier in a more intuitive way than current practice\end{tabular} &
  \begin{tabular}[c]{@{}l@{}}P2, P4, P9, P10 \\ (4)\end{tabular} \\ \hline
\multirow{4}{*}{Outcomes} &
  Model accuracy &
  \begin{tabular}[c]{@{}l@{}} Shown through behavior data\\ Shared during interview \end{tabular} &
  All (12) \\ \cline{2-4} 
 &
  \multirow{2}{*}{Model attention} &
  \begin{tabular}[c]{@{}l@{}} Shown through behavior data\\ Shared during interview\end{tabular} &
  All (12) \\ \cline{3-4} 
 &
   &
  3 particularly remarked fairness &
  P2, P5, P10 (3) \\ \cline{2-4} 
 &
  Model robustness &
  \begin{tabular}[c]{@{}l@{}}4 particularly remarked the approach can improve a model's\\robustness against malicious attacks\end{tabular} &
  \begin{tabular}[c]{@{}l@{}}P1, P2, P3, P4 \\ (4)\end{tabular} \\ \hline
\end{tabular}%
}

\caption{Study 2 user feedback regarding \system's advantages.}
\label{tab:table2_pros}
\end{table}

The behavioral data we collected shows that all participants generated the model that outperforms (1) its model accuracy, (2) the overlap between the model's focus and the relevant object types (IoU), and (3) the proportion of reasonable attention out of all images in a test set. 
The average accuracy of 12 users’ fine-tuned models (${\mathrm{M}}_{exp}$) was 82.95\%, with an average IoU of 0.39 (``Intersection over Union'' with respect to the attention ground truth of the user-defined gender-related object: ``person''), and the average proportion of reasonable attention was 89.55\% (see \RThreeSide{Item.8}\FourK{Fig.~\ref{fig:fig_merged}-A}). All these performances outperformed both the initial model (model M: accuracy = 47.6\%, IoU = 0.12, attention reasonability = 51.8\%) and the model that applied state-of-the-art fine-tuning method without attention (model ${\mathrm{M}}_{base}$: accuracy = 79.0\%, IoU = 0.26, attention reasonability = 79.4\%).

Regarding the attitudinal survey data, every absolute and relative question's mean was over 4.
In terms of absolute questions, 100\% of ratings were above 4-``neutral'' (M = 6.19, SD = 0.67).
This indicates that participants were satisfied with the overall quality of the workflow and the system.
Regarding the relative questions, 89.6\% of ratings were above 4-``neutral'' (M = 5.94, SD = 1.24), which indicates that they felt applying the workflow and the system can practically improve their current practice.



\subsubsection{[RQ1-a] \textbf{Workflow: Adjusting model attention as a CNN steering method}}

\begin{figure}[!t]
    \centering
    \includegraphics[width=0.98\textwidth]{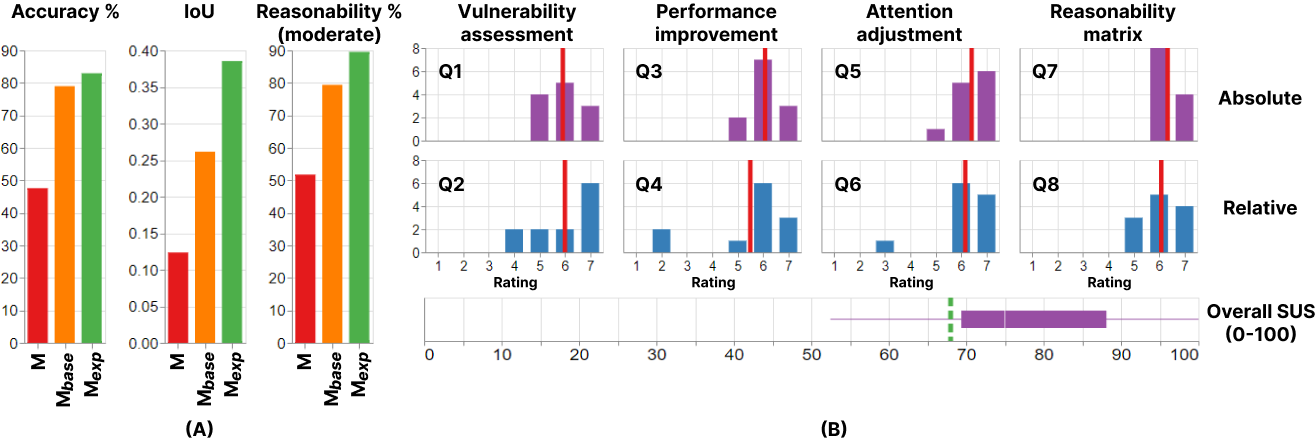}
    \caption{\FourK{}\textbf{Model improvement by \system and user feedback (Study 2):} \textbf{(A)} Model performance comparison between three model conditions: \textbf{M} (the initial classifier), ${\mathrm{\textbf{M}}}_{base}$ (the state-of-the-art fine-tuned model), and ${\mathrm{\textbf{M}}}_{exp}$ (\system's fine-tuned model using attention). Only ${\mathrm{\textbf{M}}}_{exp}$'s performance (green bars in \textbf{(A)}) is averaged from 12 models steered by our 12 participants. \textbf{(B)} User responses in two valuation perspectives (``absolute'' on the first row, and ``relative'' on the second row) from the input survey (Q1, Q2), the output survey (Q3 to Q8), and the System Usability Scale (SUS) survey (the green dotted line in the box plot is the above-average level of 68; Md = 75, M = 76.88, SD = 14.70).}
    \label{fig:fig_merged}
    \Description{Contains 2 parts, (A) is a model performance comparison between three model conditions in bar plots: the initial classifier, the state-of-the-art fine-tuned model, and DeepFuse's fine-tuned model using attention, in terms of accuracy, IoU, and reasonability. (B) has 8 bar plots and 1 box plot for all the survey responses.}
\end{figure}

After completing the user studies, the majority of users strongly agree that adjusting local explanations can effectively improve model performance (Q5 rating: M = 6.42 out of 7-``strongly agree'', SD = 0.64, as shown in Fig.~\ref{fig:fig_merged}-B). Also, people think their current modeling processes can be practically improved by considering the attention adjustment method (Q6 rating: M = 6.17 out of 7-``strongly agree'', SD = 1.07).


During interviews, all participants shared their positive impressions about the effectiveness of attention adjustment in improving model accuracy, which is the primary objective of conducting model fine-tuning. They also confirmed that the impact of contextual bias was reduced as attention quality increased by attention steering. By adding a new perspective from humans, a model also becomes fairer in making predictions for each target class (P2, P5, P10).
Participants (P1, P2, P3, P4) with experience in model attack and defense shared the possibility of using our method to improve the robustness of the models against backdoor attacks, letting the model ignore small perturbations on an image and focus on the right area. We learned that after trying our method, people gained awareness of considering human-in-the-loop and visual-based approaches in model steering since most of the ML researchers use algorithmic approaches for handling contextual bias, such as data augmentation, hyperparameter tuning, ensemble methods, etc., rather than extensively using visualization in the fine-tuning process.

\subsubsection{[RQ1-b] \textbf{Workflow: Adding quality of model attention in evaluating CNNs}}
Based on the feedback, users agree that using an attention evaluation method (e.g., reasonability matrix as \system guided, based on Gao et al.~\cite{gao2022aligning}) is effective in diagnosing model vulnerabilities (Q7 rating: M = 6.33, SD = 0.47, see Fig.~\ref{fig:fig_merged}-B), and they are very likely to use this method for improving future practices Q8 rating: M = 6.08, SD = 0.76).


Participants think that the attention assessment features in \system provide more diverse and rigorous perspectives in assessing a model's vulnerabilities, especially the reasonability matrix, which can be seen as an expansion of the accuracy dimension to understanding ``why'' a model underperforms (P1, P3, P5, P6, P8, P9, P10, P12). P1 and P4 endorsed the necessity of equipping a reasonable matrix assessment step in checking the model’s decision-making.
The matrix interpretation was straightforward to most users, as it is related to the widely-used confusion matrix concept in the data science domain.
The dynamic shifts of model vulnerability were well presented as shown by the reasonability matrix (3 vulnerable sub-groups, ``UIA - unreasonable inaccurate’’, ``UA - unreasonable accurate’’, and ``RIA - reasonable inaccurate’’).
One major task we designed for users to achieve was the recognition of a backdoor attack in the data (i.e., added green star markers which may trigger a false prediction by the model), and all participants were able to identify the impact of the attack by evaluating attention quality using the reasonability matrix.

\subsubsection{[RQ2-a] \textbf{System: How \system improved CNN diagnosis}}

After comparing with people's current practices, \system was confirmed as a useful (Q1 rating: M = 5.92 out of 7-``extremely useful'', SD = 0.76, see Fig.~\ref{fig:fig_merged}-B) and easier tool (Q2 rating: M = 6.0, SD = 1.15) in understanding model vulnerability, benefiting from the labor-efficient mechanisms.


The step-by-step nature of the assessment process in \system allows users to systematically detect both contextual and manipulated bias in the data, making it easier to reduce model vulnerability (P3, P9, P12). People believe this GUI design can significantly reduce human effort in coding and visualization management for comprehensively assessing a CNN (P2, P3, P5, P6, P7, P8, P9, P10, P12). ML engineers are well aware of the advantages of using visualization to compare metrics and surface bias, but it is a cumbersome task (e.g., repetitive file creation and loading, lack of visual-based explorers for local explanations, etc.). Instead, people mostly use command lines and unintuitive numeric comparisons for checking vulnerabilities.


One important feature that people liked was the local explanation grouping by detected objects (e.g., ``person'', ``bicycle'', etc.), which allowed them to check attention quality and accuracy changes within the common object level (P2, P3, P6, P9, P12). 
Some users pointed out that having consistent criteria for annotating attention quality regarding the classification task could be tricky with subjective uncertainty (P2, P4, P6, P9, P11). P6 mentioned that during the initial exploratory analysis of some models, users might not have good/bad attention criteria for annotating the attention. 
P10 shared an experience in exploring what objects cause contextual bias, and the biggest challenge was making a reasonable assumption at first and evaluating it over time. This challenge is critical if the annotation task is outsourced to multiple people.


\subsubsection{[RQ2-b] \textbf{System: How \system improved CNN revision outcomes}}


According to survey responses, people witnessed the highly effective capability of \system in the performance steering task (Q3 rating: M = 6.08 out of 7-``extremely effective'', SD = 0.64, see Fig.~\ref{fig:fig_merged}-B).
Regarding the same task, people found it slightly more effective than current approaches (Q4 rating: M = 5.5, SD = 1.66) as 2 users who preferred their approaches and rated 2-``less effective''.


Aligning model attention with human perceptions can effectively revise a model performance, and with \system's adjustment mechanisms (i.e., attention drawing panel and boundary suggestions, as shown in Fig.~\ref{fig:fig_ui_4_2}), people can directly embed their intention and domain knowledge into the CNN (P2, P4, P9, P10). Regarding model performance comparison, people were able to reveal the overall context of the image data and the corresponding impact on the model (accuracy and attention quality) by detected object sub-grouping of \system (P1, P2, P3, P5, P6, P8, P9, P11, P12).
An industry practitioner who worked primarily on model quality assurance mentioned that the black-box models were not usually accessible for engineers outside the core ML team, and \system had features that could be practical for them to evaluate the model performance in that situation (P11). 
In the last evaluation view of \system for record-wise attention comparison (as shown on the right of Fig.~\ref{fig:fig_ui_5}), P7 was curious about the opposite shift of attention quality (i.e., a change from ``right’’ to ``wrong’’ attention after model fine-tuning) and wanted to see some quantitative measures about it.

The IoU distribution visualization was another measure in \system that could provide a rigorous comparison between model conditions (with/without attention adjustment), revealing the positive relationship between accuracy and attention quality improvement (P2, P8, P11). As people mentioned, measuring IoU was not commonly used in classification evaluation compared to segmentation tasks, and it was typically difficult to visualize.

\subsection{Discussion}

\begin{table}[]
\resizebox{\textwidth}{!}{%
\begin{tabular}{llll}
\hline
\textbf{Themes} &
  \textbf{Sub-themes} &
  \textbf{Explanations} &
  \FourI{\textbf{PIDs \& (count)}} \\ \hline
\multirow{2}{*}{Existing issues} &
  \begin{tabular}[c]{@{}l@{}}Insufficient description \\ about reasonability matrix\end{tabular} &
  \begin{tabular}[c]{@{}l@{}}It needs more description about how to interpret it for\\ people who are unfamiliar with reasonability matrix\end{tabular} &
  \begin{tabular}[c]{@{}l@{}}P4, P7, P11 (3)\end{tabular} \\ \cline{2-4} 
 &
  \begin{tabular}[c]{@{}l@{}}Ill-formed local explanation\\ visualization\end{tabular} &
  \begin{tabular}[c]{@{}l@{}}The local explanation visualization can be improved\\ (especially the polygon mask)\end{tabular} &
  P3, P10, P12 (3) \\ \hline
\multirow{4}{*}{Future suggestions} &
  \begin{tabular}[c]{@{}l@{}}Attention adjustment can be\\ subjective\end{tabular} &
  \begin{tabular}[c]{@{}l@{}}There could be subjective uncertainty in attention adjustment\\which could be avoided by providing deterministic guidelines\\with consistent criteria to new users\end{tabular} &
  \begin{tabular}[c]{@{}l@{}}P2, P4, P6, P9, \\ P10, P11 (6)\end{tabular} \\ \cline{2-4} 
 &
  Comparison on-the-fly &
  \begin{tabular}[c]{@{}l@{}}It would be useful if the model performance could instantly\\ reflect with different adjustments\end{tabular} &
  P7, P10 (2) \\ \cline{2-4} 
 &
  \begin{tabular}[c]{@{}l@{}}Advanced adjustment\\ module\end{tabular} &
  \begin{tabular}[c]{@{}l@{}}The adjustment experience can be improved by supporting\\curve and border drawing with flexible options for\\adjusting weights of the masks other than binary masks\end{tabular} &
  \begin{tabular}[c]{@{}l@{}}P1, P3, P6, P7\\ (4)\end{tabular} \\ \cline{2-4} 
 &
  \begin{tabular}[c]{@{}l@{}}Scalable interaction for\\ bigger dataset\end{tabular} &
  \begin{tabular}[c]{@{}l@{}}It needs improved computing capabilities and design features\\for bigger datasets\end{tabular} &
  \begin{tabular}[c]{@{}l@{}}P2, P3, P4, P8,\\ P11 (5)\end{tabular} \\ \hline
\end{tabular}%
}
\caption{Study 2 user feedback regarding \system's disadvantages.}
\label{tab:table3_cons}
\end{table}

Overall, the system received acceptable usability~\cite{brooke2013sus} with an average SUS score of 76.88 (SD = 14.70, see the SUS box plot in Fig.~\ref{fig:fig_merged}-B, the rated scores (0-4) were converted to a 0-100 scale based on Brooke's SUS guide~\cite{brooke1996sus}), exceeding the average SUS level of 68. There were 10 out of 12 participants (except P3 and P5) who gave above-average SUS scores.

Although this study is not for system-level comparison, we wanted to understand the effect of our fine-tuning mechanism collected from real users. We conducted Mann-Whitney U tests to confirm the significant performance improvement after using attention.
From each of the 12 participants' results, the accuracy of our fine-tuned model using attention was significantly greater than the baseline line condition (U = 0, ${n}_{base}$ = ${n}_{exp}$ = 12, p < 0.00001). The same results apply to the IoU and attention reasonability proportion comparisons.


Through the studies, we also identified disadvantages of our system that need to be improved (as shown in Table~\ref{tab:table3_cons}). 
Regarding the interpretation of the reasonability matrix produced by users' annotation and model prediction, the guidelines can be more formally provided to be acceptable in the ML community (P4, P5, P11). The styles of attention visualization (i.e., color-scale, gray-scale, and polygon mask) need improvement, especially since the orange polygon mask was not visually clear for P3 and P10. It can be solved by having color and opacity adjustment features. 
People also raise the potential inconsistency issue in attention adjustment, where users may have subjective options and criteria about where the ``right'' attention should be. \system needs to further provide more deterministic guidelines in attention adjustment for more complex task types, especially for tasks that require domain expertise (e.g., TB diagnosis in chest X-ray images~\cite{liu2017tx}).
With this uncertainty in attention adjustment, P7 and P10 suggested an instant performance comparison feature to reflect the model improvement on the fly as people annotate, which can be a future direction in active learning to have simultaneous updates while labeling in progress~\cite{wang2016cost}.
About the attention adjustment module, people suggested that the drawing feature should be optimized for drawing curves and near image borders, as it was not easy to do so (P1, P3, P6). P5 suggested existing smart drawing features (e.g., image matting tool in Photoshop~\cite{shelly2008adobe}) to be added. P7 thinks that binary mask drawings might not be enough for the best attention guidance used in fine-tuning the model. A solution could be giving higher weights toward the centroid of the attention areas.

\TwoACSide{Item.5(b)}\OneB{With the current data size and task setting in S2, the trade-off between manual workload and model improvement may not be as significant since the overall workload was not overwhelming and considered labor-efficient compared with existing assessment methods. Though evaluating attention maps could be a labor-intensive step, diagnosing and optimizing the model's vulnerability were effective and easy to use based on users' feedback. The annotation steps were incorporated with AI-supported automation (bulk annotation, object detection, object relevance filtering, adjustment recommendation, etc.) to reduce both users' cognitive and labor workloads while gaining better performance. However, as data size increases, this labor-performance trade-off becomes essential, and more specifically, scalability solutions should be explored to reduce human labor while maintaining good fine-tuning performance. We further discussed scalability considerations regarding the trade-off in the next section (6.3).}

%% file: sections/06_IMP.tex
\section{Implications for Design Beyond XAI}

Through S1 and S2, we learned several insights from our participants.
While listening to their voice and questions, and observing the way they perceive \system after their usage, we learned that at the heart of people's pursuit of grounding their models into their practice, one of the core challenges they encounter seems to understand how they can harmonize between the way they see the CNN should suppose to work and the way CNNs actually work.
When they identify such a gap through XAI-driven tools, the upcoming challenge seemed to be to know how to reconcile such a gap efficiently and effectively. 
We reflect on this aspect of beyond XAI---how to help a user to shift their learned insights to actionable plans---and list up possible research directions that the HCI and CSCW communities can consider in designing future XAI or steerable AI tools to help practitioners ``in the trench''.

\subsection{Correlating Model Attention and Model Accuracy}

One of the overarching questions we wanted to understand was how the model attention seen as reasonable by the human mind could also result in accurate prediction. 
Perhaps that was the reason we decided to use the reasonability matrix. 
If reasonable attention and accurate prediction are aligned together, the reasonable accurate instances (i.e., accurate for the right reason) and unreasonable inaccurate instances (i.e., inaccurate for the wrong reason) should increase while the unreasonable accurate and reasonable inaccurate instances should decrease.
The tendency we saw was positive. We observed the reasonable accurate instances increased while the unreasonable accurate instances decreased from most participants. 
At least from our setting, adding more human reasoning to the model's way of thinking has increased the model's gaze toward intrinsic objects, resulting in an accuracy increment. 
However, one segment that didn't change was the reasonable inaccurate group. 
We think understanding the reason when and why the model makes inaccurate predictions despite the reasonable gaze should be closely related to improving model performance. 

Regarding research in Fairness, Accountability, and Transparency (FaccT), a dominant view is that human input or intervention may be required to realize a model that retains FaccT with the cost of model accuracy drop. 
We hope to understand the effective way to correlate the right reason, and accurate prediction can motivate the development of a fair, robust, and accurate model~\cite{hong2019disseminating, hong2020towards}.
In general, we believe it is important to understand how to align human reasoning and model accuracy.
Shao et al. argue that humans ``arguing'' against DNNs when explanations are not reasonable can benefit the model~\cite{shao2020towards}. 
A railroad cannot be a train~\cite{lee2021railroad}, a snowboard is not a man~\cite{hendricks2018women}, and a shopping cart should not be a woman~\cite{zhao2017men}.
Lastly, while human-guided ML has a potential and good cause~\cite{gil2019towards, collaris2020explainexplore}, finding a way to cut down the human-side labor is another important perspective from the two studies.

\subsection{Generalizability Consideration: \OneA{Beyond} Binary Classification}

We started to test the idea of direct steering of model attention through local explanation from the binary classification problem for reasons---simplicity of the problem and well-annotated datasets.
After using \system, several participants shared their feedback and curiosity on how our pipeline can be applied in more advanced vision-based tasks.
The design we provided in binary classification can be relatively simpler than the aforementioned cases.
As the model's task gets more complex and diverse, new designs customized to the particular task type and application area should be required to understand the generalizability of our findings.

\TwoACSide{Item.5(a)}\OneA{Methodologically, local explanation-based attention steering is not limited to binary classification tasks.
The future design can be explored to enhance CNN models for handling different tasks, such as multi-class classification, object detection, and segmentation tasks, which could possibly be expanded from processing images to videos.
The core user flow beneath \system in CNN steering is as follows:}
First, the \OneA{user flow} allows human users to define reasonable and unreasonable types of attention depending on task goals. 
Next, the \OneA{user flow} motivates reasonable attention types and penalizes unreasonable attention types in a fine-tuning process suggested in Explanation-guided Learning~\cite{gao2022going}.
Finally, the designer can provide a dashboard that helps users to understand how their indicated directions were reflected in the model revision process.

While the flow can be generally applicable, the way a designer facilitates a user's definition of reasonable and unreasonable attention type should be carefully implemented depending on the type of problem.
\OneA{For example, in a multi-class classification or object detection task for different animals, users can employ attention logic that penalizes background and motivates foreground objects to build a more reasonable and high-performing model. 
As mentioned in 5.1.1, local explanation methods can be applied to different layers of a CNN to produce different levels of granularity.
If the task goal requires a coarse granularity detection of a bounding box, applying local explanation visualization at the last layer of CNN can be suitable. However, if it needs more fine-grained granularity of closed curve for semantic segmentations, producing local explanations on both the first convolutional layer for edge-level of detail and the last convolutional layer for object-level detail can be considered, providing more depths of local explanation for users to evaluate.}
Finally, we noted P7's suggestion about extending this flow to a more advanced video level of object classification, detection, and segmentation model steering. 
Due to the data volume, special design considerations need to be applied in such a task. 
However, upon the efficient design for indicating reasonable and unreasonable attention types, we believe that it is possible to apply the suggested flow to the problem space.

\subsection{Scalability Consideration: Hundreds vs. Millions}
Despite the promising performance of the model steering method, scalability remains an essential concern raised by several participants (P2, P3, P4, P8, P11), as many real-world image classification tasks involve millions of images.
Human scalability has been a crucial issue in HCI, CSCW, and beyond---while \Author{Misc.}\FourL{the data size can easily go up to millions and trillions in training state-of-the-art models, human cognition remains flat~\cite{yan2022flatmagic}.} 
Even if we can surface millions of images to users, it may not be possible for them to scan images serially and achieve sensemaking.
Generally, to successfully devise a scalable design, we believe that the number of images users have to go over should still not exceed thousands, and the amount of time they may spend should not exceed one hour, as recent data annotation literature suggests~\cite{choi2019aila}.
Herbert Simon remarked that ``wealth of information creates a poverty of attention''~\cite{simon1981sciences}.

\begin{figure*}[!t]
    \centering
    \includegraphics[width=1.0\textwidth]{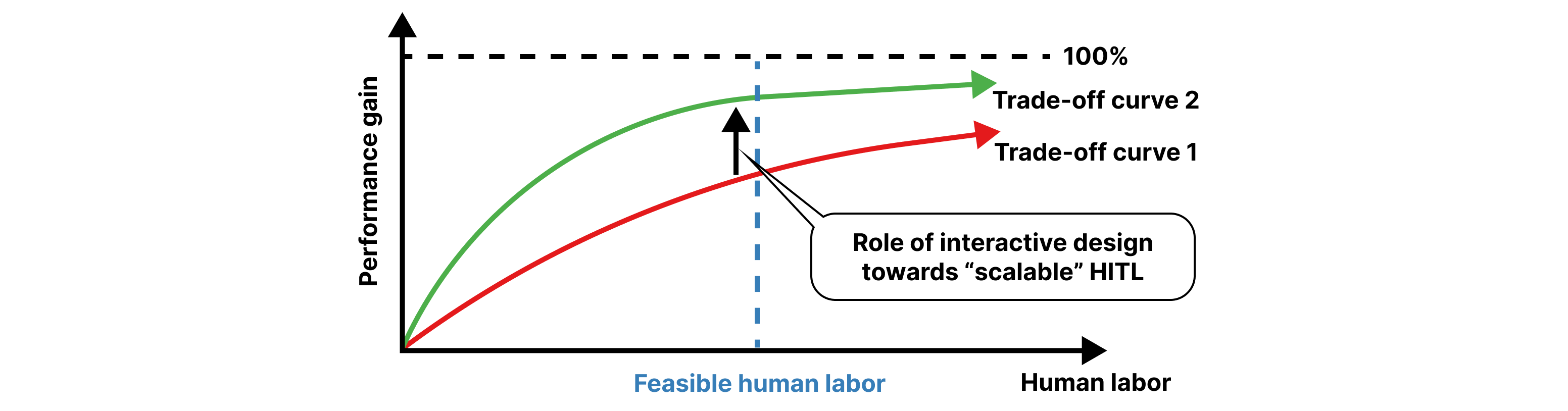}
    \caption{\OneB{An illustration of scalability optimization on a human-in-the-loop (HITL) application regarding the trade-off between human labor (as data size increases) and performance gain. The model performance can be improved faster before hitting the bottleneck of feasible human labor (the blue vertical dotted line) when the HITL processing capability is scaled up from ``trade-off curve 1'' (the red line) to ``trade-off curve 2'' (the green line).}}
    \label{fig:fig_tradeoff}
    \Description{A line chart for explaining how scalability optimization can affect the trade-off between human labor and performance gain.}
\end{figure*}

\OneB{As the trade-off between human labor and performance gain in human-in-the-loop applications is illustrated in Fig.~\ref{fig:fig_tradeoff}, when users spend more effort as data size increases, the model will gain better performance until the workload hits the bottleneck of feasible human labor. We aim to make the curve of labor-performance trade-off steeper (from ``curve 1'' to ``curve 2'' shown in Fig.~\ref{fig:fig_tradeoff}) through scalability optimization to improve the impact of human workload on performance gain. By devising ``scalable'' human-in-the-loop approaches, model performance could be further improved with the feasible amount of available human labor.}

\TwoACSide{Item.5(b)}\OneB{While \textbf{every} human-in-the-loop approach can suffer the bottleneck of limited information, labor source, session time, etc., ultimate breakthroughs in human-in-the-loop and interactive ML designs could come from scalability strategies.}
We introduce how some of the design strategies can be adopted in the design space of Beyond XAI. 
First, one can consider sampling from the whole dataset. 
Modern computer vision models can yield keywords of objects and context in the scene. Using such additional information extracted from the vast dataset, it is possible to define major and minor clusters of images. The new design may help users proceed with a small portion of sampled images derived from such clusters to reason the whole dataset and typify reasonable and unreasonable attention types accordingly.
Second, one can consider examining images based on the sequence built from Active Learning, \Author{Misc.}\FourL{a technique that chooses the fewest unlabeled data possible that could maximize the model accuracy gain~\cite{chung2021understanding}}.
Applying active learning techniques is common in data annotation research, which can help reduce the required size of images to reason.
Third, devising further intelligent features that can automate the current workflow can facilitate the process as well. 
Some features that need manual investigation can be automated in future designs.
Finally, if there is a strong rationale for investing more human resources, one can consider crowdsourcing.

\subsection{Data Iteration and Continual Lifelong Learning}

\system's capability of figuring out the vulnerability through local explanation is closely related to the capability of fortifying the dataset by adding more examples that can remove the contextual bias.
Such ``data iteration'' is not uncommon in practice. 
To improve the model, the most fundamental way is to improve data. For instance, \textit{Chameleon} lets users compare data features, training/testing splits, and performance across data versions~\cite{hohman2020understanding}. 
When combining the data iteration with model steering using local explanations, one could derive some interesting design ideas that can help ML engineers to better find, search, and add the dataset.

While improving the model with new data can be straightforward, a few issues need to be considered when steering models through local explanations. 
First, it is necessary to understand what learning strategy can be more effective between the case where stacking every dataset in one place and retraining the model and the case of iteratively adding the new dataset and making the model ``evolve'', 
In general, the first case can yield a high-performing model than the second case due to the chance of catastrophic forgetting, which is a problematic and almost inevitable drawback~~\cite{kirkpatrick2017overcoming}. 
In recent years, the concept of continual lifelong learning has emerged~\cite{parisi2019continual} and provided a breakthrough. 
Understanding which strategy can yield what strengths and weaknesses in the scenario of data iteration with local explanation reasoning would be necessary.

\subsection{Improving Fine-Tuning}
This work is the first study that observes how ML engineers experience techniques in the Explanation-guided Learning framework in fine-tuning their model and perceiving the difference.
While we saw participants satisfied with the progress they made with the RES framework, we introduced a few directions on how the RES framework can be evolved to design an improved model steering environment in the future.

One important direction is how to design a better quantitative measurement to assess the quality of the steered attention during the fine-tuning process. 
Simple distance-based metrics such as Mean Squared Error (MSE) or Intersection over Union (IoU) scores that are calculated purely based on the alignment of each feature can hardly comprehensively reflect the quality of the adjusted attention, as they completely ignore the correlations among visual features.
One potential remedy to this issue is also to leverage fidelity-based metrics, which aim at evaluating how faithful the model's attention is with respect to the model's prediction.
The assumption behind this is that the `right' attention should contain sufficient information for the model also to make the `right' prediction \cite{deyoung2019eraser, mitsuhara2019embedding, situ2021learning}; while on the other hand, removing the attention should also lead to significant negative impact for the model to make the correct prediction \cite{deyoung2019eraser, mitsuhara2019embedding, ismail2021improving}.
However, it is still not clear and challenging to propose a single metric that can together measure the faithfulness and the degree of alignment with the human annotation to make a more comprehensive assessment of the attention quality.

Another possible topic is how to leverage multiple annotations from different users for a single sample~\cite{chung2019efficient, chung2021understanding}. 
As obtaining more than one annotation can be helpful to boost the reliability of the human boundary for attention adjustment, it poses challenges on how to align model attention with multiple ground truth boundaries. 
While a simple way out can be using the 50\% consensus or majority vote over all the available annotations, useful information can be lost during the aggregation. Thus, new techniques are in demand to leverage each annotation effectively.

%% file: sections/07_Conclusion.tex
\section{Conclusion}

In this work, we examined our inquiry of how we can design a direct feedback loop between a human and a CNN through local explanations. 
In particular, we designed and developed the first interactive system to help a user adjust the local explanation results regarding the gaze of CNNs.
We applied our interactive design in the problem space of contextual bias for CNN engineers. 
With the S1, we learned ML engineers' practical challenges and desires, converting the insights to design considerations that could improve how we use local explanations in model diagnosis and steering.
With \system, we conducted S2 and found how \system can provide a better workflow and experience to CNN engineers.
At the same time, we also found limitations and future research directions.
In particular, we boiled down and shared in Implications for Design beyond XAI within the categories of (1) correlating model attention and model accuracy, (2) generalizability consideration, (3) scalability consideration, (4) data iteration and lifelong learning, and (5) improving fine-tuning. 
We hope this work can benefit researchers and practitioners who seek to understand how to make XAI-driven insights actionable in steering AI.

%% file: sections/98_Appendix.tex
\appendix
\FourC{\section{Study 1 interview questions}}
\TwoACSide{Item.3}

\subsection*{About you}
    \begin{itemize}
        \item Can you explain your role in your company?
    \end{itemize}

\subsection*{Your models and development settings}
    \begin{itemize}
        \item Can you explain the purpose, input, and output of your models for which you used model saliency/attention? 
        \item Can you walk us through your process of building your model? E.g., how to collect the training set, how to train your model, how to improve your model performance, how to debug?
    \end{itemize}

\subsection*{Use of saliency maps}
    \begin{itemize}
        \item Can you explain the way you use saliency maps in understanding your model’s behavior?
        \item Can you explain the way you use saliency maps in supervising/improving your model’s behavior?
    \end{itemize}

\subsection*{Working on fair/robust/accurate models}
    \begin{itemize}
        \item Can you explain your experience/effort towards building more fair DNN models?
        \item Can you explain if attention/saliency was useful or not? 
    \end{itemize}

\subsection*{Your tools, challenge, and wish list in the future}
    \begin{itemize}
        \item Can you explain the types of tools that you use for understanding/improving your DNN models?
        \item Can you explain the challenges you experience while interacting with your DNN?
        \item What new tools/features do you wish to have in the near future to make your life better?
    \end{itemize}







\ThreeB{\section{Study 2 System Usability Scale (SUS) survey~\cite{brooke1996sus}}}
\RThreeSide{Item.5}

\subsection*{Indicate your degree of agreement for each of the 10 statements (on a Likert scale from 1-``strongly disagree'' to 5-``strongly agree'')}
    \begin{enumerate}[label=\arabic*.]
        \item I think that I would like to use this system frequently.
        \item I found the system unnecessarily complex.
        \item I thought the system was easy to use.
        \item I think that I would need the support of a technical person to be able to use this system.
        \item I found the various functions in this system were well integrated.
        \item I thought there was too much inconsistency in this system.
        \item I would imagine that most people would learn to use this system very quickly.
        \item I found the system very cumbersome to use.
        \item I felt very confident using the system.
        \item  I needed to learn a lot of things before I could get going with this system.
    \end{enumerate}
